\documentstyle[12pt,epsfig]{article}
\begin{document}

\title{The Quantum Spherical $p$-Spin-Glass Model}
\vskip 10pt
\author{
Leticia F Cugliandolo$^{1,2}$, D. R. Grempel$^3$ and 
\\
Constantino A da Silva Santos$^2$
\\
$^1$Laboratoire de Physique Th{\'e}orique de l'{\'E}cole Normale 
Sup{\'e}rieure, 
\\
24 rue Lhomond, 75231 Paris Cedex 05, France
\\
$^2$Laboratoire de Physique Th{\'e}orique  et Hautes {\'E}nergies, Jussieu, \\ 
5{\`e}me {\'e}tage,  Tour 24, 4 Place Jussieu, 75252 Paris Cedex 05, France
\\
$^3$CEA-Service de Physique de l'{\'E}tat Condens{\'e}, CEA-Saclay,
\\
91191 Gif-sur-Yvette CEDEX, France
}

\date\today
\maketitle

\begin{abstract}
We study a quantum extension of the spherical $p$-spin-glass
model using the imaginary-time replica formalism. We solve the model
numerically and we discuss two analytical approximation schemes that
capture most of the features of the solution.
The  phase diagram and the physical properties of the
system  are determined in two ways: by imposing the usual conditions of
thermodynamic equilibrium and by using the condition of 
marginal stability. In both cases, the phase diagram 
consists of two qualitatively different 
regions. If the transition temperature is  
higher than a   
critical value
$T^{\star}$, quantum effects are qualitatively irrelevant
  and the phase transition is {\it second} 
order, as in the classical case. However, when quantum
fluctuations depress the transition temperature below $T^{\star}$, 
the transition becomes {\it first order}. The susceptibility is
discontinuous and shows hysteresis across the first order line, 
a behavior reminiscent of that observed in  
the dipolar Ising spin-glass LiHo$_x$Y$_{1-x}$F$_4$ in an external 
transverse 
magnetic field. We discuss in detail the thermodynamics and the
stationary dynamics of both states. The spectrum of magnetic
excitations of the equilibrium spin-glass state is gaped, leading to
an exponentially small specific heat at low temperatures. That of the
marginally stable state is gapless and its specific heat varies
linearly with temperature, as generally observed in glasses at low
temperature.
 We show that 
the properties of the marginally stable state 
are closely related to those obtained in studies of the real-time 
dynamics of the system weakly coupled to a quantum thermal bath.
Finally, we discuss a possible application of our results to 
the problem of polymers in random media.

\end{abstract}
\newpage

\section{Introduction}
\setcounter{equation}{0}
\renewcommand{\theequation}{\thesection.\arabic{equation}}

The description of spin glasses in terms of classical statistical 
mechanics is generally  
justified since in most cases the transition 
temperature $T_g$ is too high for quantum effects to
be relevant. 
In some cases of practical interest, however, 
 quantum fluctuations, controlled by an external parameter
 ({\it e.g.} magnetic field, doping, pressure), may reduce $T_g$ down 
to arbitrarily low values and even suppress the glass transition
altogether if they are strong enough.
Among the experimental systems belonging to this class we may cite 
magnetic systems such as 
LiHo$_x$Y$_{1-x}$F$_4$\cite{aeppli1}, La$_{2-x}$Sr$_x$CuO$_4$
\cite{shirane} and UCu$_{5-x}$Pd$_x$ \cite{vollmer} as well as some 
randomly mixed hydrogen-bonded 
ferro-antiferroelectric crystals \cite{exp2}.
The theory of phase transitions in systems such as these must
necessarily
 take 
into account their quantum
mechanical nature. 

A question of fundamental interest is whether there exists any {\it
qualitative} 
differences between quantum spin-glass systems  
and their classical counterparts. 
This issue has been extensively investigated experimentally
\cite{aeppli1} in the case of the compound LiHo$_x$Y$_{1-x}$F$_4$. 
This is a site-diluted
derivative of the  dipolar Ising ferromagnet $\rm LiHoF_4$. 
For $x<1$, the positional 
disorder of the magnetic $\rm Ho^{3+}$ ions makes the long range dipolar
couplings 
random. LiHo$_x$Y$_{1-x}$F$_4$ is thus 
a spin-glass with a freezing
 temperature $T_g(x)$ \cite{aeppli1}. 
The application of an external magnetic field $H$   
{\it transverse} to the easy axis allows quantum tunneling  
through the barrier separating the two degenerate ground states
 of the $\rm Ho^{3+}$ ions. Quantum tunneling competes against
spin freezing and the spin-glass ground state is expected to be
destroyed at all temperatures if the tunneling frequency  ($\propto
H^2$) is sufficiently high. 
It was found experimentally that
 $\rm LiHo_{0.167}Y_{0.833}F_4$ is paramagnetic 
at all
temperatures 
above $H_{c}(0) \approx 12$ kOe \cite{aeppli1}. Below this critical
field,  paramagnetic ({\sc pm}) and spin-glass ({\sc sg}) 
phases exist, separated by a line $H_{c}(T)$. 
Above 25 mK, the  phase transition 
is second-order and signaled  
by a divergence of the non-linear susceptibility $\chi_3$. However, 
below 25 mK, the divergence of $\chi_3$ at the transition 
becomes  a flat maximum at the position of which
 the imaginary part of the low-frequency linear
susceptibility $\chi''(\omega)$ has a jump. These features strongly
suggest that, at low transition temperatures [and in particular at
the quantum critical point at ($H=H_{c}(0), T=0$)] the
 transverse-field-induced spin-glass 
transition becomes  
{\it first order} \cite{aeppli1}. 
This conclusion has recently received further support from the
observation of hysteresis in the linear susceptibility measured as a
function of the transverse field \cite{aeppli2}.

From the theoretical point of view, there is no evidence 
for this type of transition 
in the standard models for quantum spin glasses discussed in the literature. 
The quantum Edwards-Anderson (EA) \cite{finite_D} and Sherrington-Kirkpatrick
(SK) \cite{bm-quantum,SK-transv,large_D} models are known to undergo
 second order
transitions as also do models of metallic
spin glasses \cite{metallic}. 
Interestingly enough, this type of scenario 
does occur in less standard models with multi-spin
interactions \cite{Gold,Niri,GS,Opper,Cugrsa}. 
 
In this paper we study in detail the properties of
one of these models, the
$p$-spin spherical model with random interactions. We solve it 
 using two different
approaches.
The first one consists in imposing the usual 
conditions of  equilibrium to find the possible thermodynamic states of
the system. In the second one, these are determined by imposing a
condition of marginal stability.
In both cases the problem is solved using a 1-step replica symmetry
breaking ({\sc rsb})
{\it Ansatz} shown to be exact just as it is in the classical case.  
We find that, in  terms of a quantum 
parameter $\Gamma$ to be defined below, there is a line $\Gamma_{c}(T)$ in the
$\Gamma-T$ plane that separates {\sc sg} and {\sc pm} 
phases. This line ends at a quantum critical point at
$T=0,\,\Gamma=\Gamma_{c}(0)$ 
above which the system is paramagnetic at all
temperatures. 
$\Gamma$ thus plays a role similar to that of the transverse field in
the experiments described above. One of the main results of this paper
is that a 
tricritical point $(T^{\star},\Gamma^{\star})$ divides the
transition line in two sections. For $T \ge T^{\star}$, the SG
transition is second order and the behavior of the 
quantum system is in all respects similar to that of the classical
one. For $T < T^{\star}$, quantum
fluctuations drive the transition first order. There is latent heat
and the magnetic
susceptibility is discontinuous and shows 
hysteresis across the first-order line. This is reminiscent of the
 behavior observed in LiHo$_x$Y$_{1-x}$F$_4$ in a transverse magnetic
field although, as it will be seen, important differences between 
the predictions of this model
and experiment exist.

It should be noted that other quantum \cite{Gold,Niri}
and classical \cite{GS,Opper}  models with similar 
characteristics were previously discussed in the literature but the
connection with the experimental results was apparently not realized. 
 
Several subtle points arise in 
this and related models \cite{Niri,GS} when one has to choose
 between several possible equilibrium or marginally
stable solutions. They stem from the existence of multiple  
stable {\sc pm} and {\sc sg} solutions in finite regions of phase space. 
To choose between them one has resort to physical arguments that
 go beyond the usual prescriptions of replica theory.
We first show how this problem arises within an improved
``static approximation'' \cite{bm-quantum} that, despite its
simplicity, contains the essential
physics of the model. We then discuss it in the framework of 
the exact numerical solution of the model.

We investigate the {\it stationary} 
dynamics of the equilibrium and mar\-gi\-nal\-ly
stable {\sc sg} states and show that they are quite different. 
We find that the magnetic
excitation spectrum of the spin-glass state is gaped, leading to
an exponentially small specific heat at low temperatures.  The
spectrum  of the
marginally stable state is gapless and its specific heat varies
linearly with temperature. A power-law behavior of $C_v(T)$ is
commonly observed in glasses at low-temperatures~\cite{Zeller,Osheroff} 
and is explained with
models based on a distribution of two-level systems~\cite{twolevel}. 
In this approach there are no such two-level systems but a linear $C_v(T)$
stems from the fact that the condition of marginal stability selects 
 flat directions in phase space as 
opposed to the equilibrium condition that selects well-defined minima 
of the free-energy.

We show explicitly that the use of the condition of marginal stability in 
the Matsubara replica approach allows us to obtain, from a purely
static calculation, partial 
information about the {\it non-equilibrium} real-time 
dynamics of the same system in contact with 
an environment, in  the 
long-time, weak-coupling limits (taken in this
order). We show 
that the transition line
coincides with the dynamic transition line determined from the real-time
formalism. We find that it also changes from second to first 
order at a tricritical point. This feature was also found in a study
of the real-time dynamics of the model coupled to a bath where it is signaled by 
a jump in the asymptotic 
energy density when the system is driven across the transition line~\cite{Homero}.
We show 
 that  $m$, the break point in Parisi's {\sc rsb} scheme, 
coincides with $T/T_{\sc eff}$
\cite{leti1}, where $T$ is the temperature of the 
quantum environment (that may be zero) and $T_{\sc eff}$ is the 
dynamically generated effective temperature \cite{Cukupe}.
Finally, we demonstrate that the time-dependent correlation function
calculated in the replica approach  with 
the condition of marginal stability coincides, in the long
waiting-time and weak-coupling limits, with the outcome of the dynamic
calculation for the stationary part of the symmetrized 
correlation function~\cite{footnote}. As in the classical problem, 
equilibrium and marginal 
results can be interpreted in terms of the solutions of the TAP
equations extended to include quantum 
fluctuations~\cite{Bicu}.

The paper is organized as follows. In Section \ref{themodel}
we present the model and derive its free-energy density with 
the Matsubara replica formalism. We discuss in detail 
several possible interpretations of the model and its relation 
to other ones that have already been studied. In Section \ref{static_approx}
we introduce a refined static
approximation which enables us to find the correct qualitative behavior 
of the system in the whole phase space. We analyze
 the consequences of using the static and marginal 
prescriptions for the spin-glass phase  within this 
approximation. 
In Section~\ref{exact} we present the exact numerical 
solution of the model distinguish again between the equilibrium and marginal
cases. We also present a low-temperature, low-frequency  
approximate solution, that goes beyond the static 
approximation, and yields results for the real-frequency dependence of equilibrium correlation functions.
The numerical results are presented in Section~\ref{numerical}.
The connection between our results and the real-time dynamics of the
system is 
established in Section~\ref{real_time}
where we compare the results for several quantities obtained in the
static and dynamic formalisms. Finally, in Section~\ref{conclusions} we draw our conclusions
and we reinterpret our results for a classical polymer in a random media.
We also briefly discuss related work in progress. 
A short account of some of our results appeared in Ref.~\cite{Cugrsa}.

\section{The model}
\setcounter{equation}{0}
\renewcommand{\theequation}{\thesection.\arabic{equation}}
\label{themodel}

We study a quantum extension of the classical  spherical 
$p$-spin-glass model \cite{crisanti} in which we reinterpret 
the continuous spins, $s_i$, as coordinate operators 
and introduce canonically conjugate 
momentum operators, $\pi_i$. Coordinate and momentum operators 
verify the usual commutation relations
\begin{equation}
[s_i, s_j] = [\pi_i,\pi_j]=0 \;  \;, \;\;\;\;\;\;\;\;
[\pi_i,s_j] = - i \hbar \delta_{ij}
\; .
\end{equation}  
The quantum spherical model is then defined by adding to the 
usual potential energy a ``kinetic energy'' term. The Hamiltonian 
reads 
\begin{equation}
H[\vec \pi, \vec s,J]= {\pi^2\over 2M} + \sum^{N}_{i_1<...<i_p} J_{i_1...i_p}
s_{i_1} ... s_{i_p} \; .
\label{eq:action}
\end{equation}
We denote $\pi^2=\vec \pi \cdot   \vec \pi$, $s^2=\vec s \cdot   \vec s$
with $\vec \pi=(\pi_1,\dots,\pi_N)$ and $\vec s=(s_1,\dots,s_N)$. 
A Lagrange multiplier $z$ enforces the mean spherical constraint
\begin{equation}
{1\over N}\sum^{N}_{i=1} \langle s_i^2 \rangle = 1 
\label{eq:lm}
\end{equation}
where the angular brackets denote the thermodynamic average.  

The interaction strengths $J_{i_1...i_p}$ are taken from a Gaussian distribution 
with zero mean and variance
\begin{equation}
\overline{\left(J_{i_1...i_p}\right)^2} = {{\tilde{J}}^2 p!\over 2N^{p-1}}
\; .
\end{equation}
Hereafter the overline represents the average over disorder. 
The second term in the Hamiltonian is a random Gaussian potential energy
with zero average and correlation
\begin{equation}
\overline{V(\vec s) V(\vec s')} = \frac{{\tilde{J}}^2N}{2} \left(\frac{\vec s \cdot \vec s'}{N}\right)^p
\; .
\label{corr_pot}
\end{equation}

\subsection{Replica formalism}

In order to study the static properties of the model we need to compute
the disorder averaged free-energy density.
The replica trick allows us to compute it as
\begin{equation}
\beta f=-{1\over N}\overline{\ln Z}= 
-{1\over N}\lim_{n_\to 0}\frac{\overline{Z^n}-1}{n} , 
\label{eq:fe}
\end{equation}
where the partition function, $Z$, is given by
\begin{equation}
Z=\mbox{Tr}\,\, e^{-\beta H}.
\end{equation}
In the Matsubara formalism, 
the disorder averaged replicated partition function
can be written as a functional integral over 
replicated periodic functions of imaginary time, $\vec s_a(\tau)$,
with $a=1,\dots,n$ being a replica index. These functions satisfy  
${\vec s}_a(\beta\hbar)={\vec s}_a(0)$. 
In order to decouple the $p$-interactions in the potential energy
we introduce in $\overline{Z^n}$ the identity 
\begin{eqnarray*}
& & 1 \propto \int D{\bf Q}\;  
\delta\left(N Q_{ab}(\tau,\tau') - {\vec s}_a(\tau) \cdot   {\vec s}_a(\tau') \right)
\nonumber\\
& &
\propto
\int  D{\bf Q} D\mbox{\boldmath$\lambda$}\; 
\exp\left[\frac{i}{2\hbar} \sum_{ab} \int_0^{\beta\hbar} d\tau 
\int_0^{\beta\hbar} d\tau'
\lambda_{ab}(\tau,\tau')
\left(
N Q_{ab}(\tau,\tau') - {\vec s}_a(\tau) \cdot  {\vec s}_b(\tau')
\right) \right].
\end{eqnarray*}
We shall hereafter use  boldface to  denote matrices in replica space.
The averaged replicated partition function can be recast as 
\begin{eqnarray}
\overline{Z^n}=\int D{\vec s} D\mbox{\boldmath$\lambda$}
D{\bf Q}\; \exp\left(-{1\over \hbar}S_{\sc eff}\right)
\; ,
\label{eq:pf}
\end{eqnarray}
with
\begin{eqnarray}
-{1\over \hbar}\, S_{\sc eff}& = &-{1\over 2}\sum_{ab}\int_0^{\beta\hbar}
d\tau \int_0^{\beta\hbar} d\tau' 
\; 
{\vec s}_a (\tau) \cdot \left[{i\over \hbar}
O_{ab}(\tau-\tau')+{i\over \hbar}\lambda_{ab}(\tau,\tau')
\right] {\vec s}_b (\tau')
\nonumber\\
& &
+{iN\over{2\hbar}}\sum_{ab}
\int_0^{\beta\hbar} d\tau
\int_0^{\beta\hbar} d\tau' \; 
\lambda_{ab}(\tau,\tau') Q_{ab}(\tau,\tau')
+ \frac{N \beta}{2} \sum_a z_a
\nonumber\\
& & 
+{{\tilde{J}}^2 N\over 4\hbar^2}
\sum_{ab}
\int_0^{\beta\hbar} d\tau \int_0^{\beta\hbar} d\tau'\; 
Q_{ab}^{\bullet p} (\tau,\tau')
\; .
\label{eq:seff}
\end{eqnarray}
We denote with a bullet the usual product: $Q_{ab}^{\bullet \, p}(\tau) =
Q_{ab}(\tau) \cdots Q_{ab}(\tau)$, $p$ times,  to distinguish it from the
operational product. $O_{ab}$ is a short-hand notation for the differential 
operator
\begin{equation} 
O_{ab}(\tau-\tau')  = i \delta_{ab} \delta(\tau-\tau')
\left( M \frac{\partial^2}{\partial\tau^2} - z_a \right)
\; .
\end{equation}
It follows that, at the saddle-point,  
the expectation value of the 
order-parameter $Q_{ab}(\tau,\tau')$ is given by
\begin{equation}
Q_{ab}(\tau,\tau') = \frac{1}{N} 
\overline{\langle {\vec s}_a(\tau) \cdot {\vec s}_b(\tau') \rangle }
\; ,  
\end{equation}
with $Q_{ab}(\tau,\tau')$ periodic in $\tau$ and $\tau'$ with period
$\beta \hbar$. 
The mean spherical constraint reads
$Q_{aa}(\tau,\tau)=1$ for all $\tau$.

Since we are studying an equilibrium problem,
all correlation functions are time-translational 
invariant. They are also symmetric in imaginary time due to the 
 time-reversal invariance of the Hamiltonian:
\begin{equation}
Q_{ab}(\tau,\tau') = Q_{ab}(\tau-\tau')= Q_{ab}(\tau'-\tau) \; , \;\;\;
\lambda_{ab}(\tau,\tau') = \lambda_{ab}(\tau-\tau')=\lambda_{ab}(\tau'-\tau)
\;.
\end{equation}
Using these properties 
we can simplify the effective action and write it as 
\begin{eqnarray}
& & 
-{1\over \hbar}\, S_{\sc eff}= -{1\over 2}\sum_{ab}\int_0^{\beta\hbar}
d\tau \int_0^{\beta\hbar} d\tau' 
\; 
{\vec s}_a (\tau) \cdot   \left[{i\over \hbar}
O_{ab}(\tau-\tau')+{i\over \hbar}\lambda_{ab}(\tau-\tau')
\right]  {\vec s}_b (\tau')\;\;\;
\nonumber\\
& &
\;\;\;\;\;\;\; +{iN\beta \over 2}\sum_{ab}
\int_0^{\beta\hbar} d\tau \, 
\lambda_{ab}(\tau) Q_{ab}(\tau)
+{{\tilde{J}}^2 N\beta \over 4\hbar}
\sum_{ab}
\int_0^{\beta\hbar} d\tau  \, 
Q_{ab}^{\bullet p} (\tau)
+ \frac{Nn\beta}{2} z
\; ,
\label{eq:seff1}
\end{eqnarray}
where we have further assumed that 
$z_a$ does not depend on the replica index. 
In the following, we work with the 
Fourier transforms
\begin{eqnarray}
\tilde s_i(\omega_k)&=& 
\frac{1}{\sqrt{\beta\hbar}} 
\int_0^{\beta\hbar}  d\tau e^{i\omega_k \tau}
s_i(\tau)
\; ,
\nonumber\\
s_i(\tau)&=& 
\frac{1}{\sqrt{\beta\hbar}} 
\, \sum_{k} e^{-i\omega_k \tau} \tilde g_i(\omega_k)
\; ,
\end{eqnarray}
with  the Matsubara frequencies given by
\begin{equation}
\omega_k = \frac{2\pi k}{\beta\hbar} \;\;\;\;\; k=0,{\pm 1},\dots
\; .
\end{equation}
This implies 
\begin{eqnarray}
\tilde Q_{ab}(\omega_k) &=& \int_0^{\beta\hbar} d\tau \, exp(i\omega_k \tau) \,
Q_{ab}(\tau)
\; ,
\\
Q_{ab}(\tau) &=& 
(\beta\hbar)^{-1} \sum_k exp(-i\omega_k \tau) \, \tilde Q_{ab}(\omega_k)
\; .
\end{eqnarray}

In terms of the Fourier transformed variables the effective action 
reads
\begin{eqnarray}
& & -{1\over \hbar}S_{\sc eff}=-{1\over 2}\sum_k
\sum_{ab}{\tilde{\vec s}}_a(-\omega_k) \cdot  
\left[{i\over \hbar} {\tilde{O}}_{ab}(\omega_k)
+{i\over\hbar}{\tilde{\lambda}}_{ab}
(\omega_k)\right] {\tilde{\vec s}}_b (\omega_k)
+ \frac{Nn\beta}{2}z
\nonumber\\
& & 
+{iN\over 2\hbar}\sum_k\sum_{ab}{\tilde{\lambda}}_{ab}
(\omega_k){\tilde{Q}}_{ab}(\omega_k)
+
{{\tilde{J}}^2 N\beta\over 4\hbar}\sum_{ab}
\int_0^{\beta\hbar} d\tau \left(
{1\over \hbar\beta}\sum_k\exp(-i\omega_k\tau)
{\tilde{Q}}_{ab}(\omega_k)\right)^p 
\nonumber\\
\label{eq:sefft}
\end{eqnarray}
where we used $\tilde Q_{ab}(\omega_k)=\tilde Q_{ab}(-\omega_k)$. 
The functional integration over the  functions $\tilde{\vec{s}}(\omega_k)$
is quadratic and can be explicitly performed. This amounts to replace 
the quadratic term in the action by
\begin{equation}
-{N\over 2} \sum_k \mbox{Tr} \ln \left[
{i \beta}\left({\tilde{\bf O} (\omega_k)}+
{\tilde{\mbox{\boldmath $\lambda$}} (\omega_k)}\right)\right].
\end{equation}
where we took into account a factor $(\beta\hbar)^{-n/2}$ that comes
from the change in variables 
$\vec s_a(\tau) \to \tilde{\vec s}_a(\omega_k)$ in the partition function.
The trace is to be taken over replica indices.

The effective action is now proportional to $N$ and, 
in the large $N$ limit, this allows us to evaluate 
the replicated partition function by the steepest descent method.
The  saddle-point equation with respect to
${\tilde{\lambda}}_{ab}(\omega_k)$
reads
\begin{equation}
\frac{i}{\hbar} \tilde {\bf Q} = 
\left(  \tilde{\bf O} +\tilde {\mbox{\boldmath $\lambda$}} \right)^{-1}
\; .
\end{equation}
By replacing this  value of 
$\tilde \lambda_{ab}(\omega_k)$ in (\ref{eq:sefft})
one obtains
\begin{eqnarray}
-{1\over \hbar}S_{\sc eff}
&=&
{N\over 2} \sum_{k} \mbox{Tr} \ln 
\left( (\beta\hbar)^{-1} \tilde {\bf Q} \right)
+{N\over 2} \sum_k \left( n - \frac{i}{\hbar}
\sum_{ab} \tilde O_{ab}(\omega_k) \tilde Q_{ab}(\omega_k) \right)
\nonumber \\
& &
+
{{\tilde{J}}^2 N\beta\over 4\hbar}\sum_{ab}
\int_0^{\beta\hbar} d\tau \left(
{1\over \hbar\beta}\sum_k\exp(-i\omega_k\tau)
{\tilde{Q}}_{ab}(\omega_k)\right)^p
+ \frac{Nn\beta}{2}z
\; .
\label{eq:sefftt}
\end{eqnarray}
Finally, the averaged replicated partition function as a function
of $\tilde {\bf Q}$ becomes
\begin{equation}
\overline{Z^n}=\exp\left(-nNG_0 \right) 
\end{equation}
where
\begin{eqnarray}
2G_0 
& = &
- \frac{1}{n} \sum_k \mbox{Tr} \ln
\left( (\beta\hbar)^{-1} {\tilde{\bf Q}} \right)
-\sum_k\left(1-\frac{i}{n\hbar}\sum_{ab}{\tilde{O}}_{ab}
(\omega_k){\tilde{Q}}_{ab}(\omega_k)\right)
\nonumber\\
& &
-{{\tilde{J}}^2 \beta\over 2\hbar n}\sum_{ab}
\int_0^{\beta\hbar} d\tau \left(
{1\over \hbar\beta}\sum_k\exp(-i\omega_k\tau)
{\tilde{Q}}_{ab}(\omega_k)\right)^p
- \beta z
\label{eq:go}
\end{eqnarray}
and the free-energy per spin is
\begin{equation}
\beta f=\lim_{n\to 0} G_0
\; .
\label{eq:fe1}
\end{equation}

The saddle-point equation with respect to  the order parameter $\tilde
Q_{ab}(\omega_k)$ reads
\begin{equation}
{1\over \hbar}(M\omega^2_n + z)\delta_{ab} = \left(\tilde{\bf Q}^{-1}\right)
_{ab}(\omega_k) + {{\tilde{J}}^2 p\over 2\hbar^2} 
\int_0^{\hbar\beta}d\tau\exp(i\omega_k\tau)Q^{\bullet p-1}_{ab}(\tau),
\label{eq:ft}
\end{equation}
that transforming back to imaginary time becomes
\begin{equation}
-{1\over \hbar}\left(M{\partial^2\over \partial\tau^2} - z\right)
\delta_{ab}\delta(\tau) = Q^{-1}_{ab}(\tau) + {{\tilde{J}}^2 p
\over 2\hbar^2} Q^{\bullet \, p-1}_{ab}(\tau)
\; .
\label{eq:motion}
\end{equation}
Equation (\ref{eq:motion})  together with the spherical constraint 
\begin{equation}
Q_{aa}(0)=1, \;\;\;\;\;\;\;\;\;\; \frac{1}{\beta\hbar} 
\sum_k \tilde Q_{aa}(\omega_k) = 1
\; ,
\label{eq:sc}
\end{equation}
are the equations that characterize the 
different phases of the model. 

From here on, we shall work with dimensionless quantities. 
We take $\tilde{J}$ as the unit of energy and $\hbar/\tilde{J}$ 
as the unit of time. Hence, we redefine the imaginary time and 
Matsubara frequencies as
\begin{equation}
\hat\tau \equiv \frac{\tilde J \tau}{\hbar} \;\;\Rightarrow  \;\;
\hat\omega_k  \equiv \frac{\hbar\omega_k}{\tilde J}
\; .
\end{equation}
The Lagrange multiplier is now given by $\hat z = z/\tilde J$ and, 
consequently, Eq.~(\ref{eq:motion}) becomes
\begin{equation}
\left( -{1\over \Gamma} {\partial^2\over \partial\tau^2} + z\right)
Q_{ab}(\tau) = \delta_{ab}(\tau) + \sum_c \int_0^\beta d\tau' \, 
{p\over 2} Q^{\bullet \, p-1}_{ac}(\tau-\tau')Q_{cb}(\tau')
\; ,
\label{eq:motion2}
\end{equation}
where $\Gamma=\hbar^2/(M\tilde{J})$ and we  have eliminated
 all hats in order to simplify the notation. 

The matrix elements $Q_{ab}(\tau)$ are the order parameters of the
model. The diagonal components $Q_{aa}(\tau)=N^{-1}\overline{\langle 
\vec s_a(\tau) \cdot \vec s_a(0) \rangle } $ in Eq.~(\ref{eq:motion2}) have no
classical analog. Indeed, in
the classical limit $0\le \tau \le \hbar \beta$ and 
$\lim_{\hbar \to 0} Q_{aa}(\tau)=Q_{aa}(0)\equiv 1$ on account of the
constraint (\ref{eq:lm}). In the quantum mechanical case 
$Q_{aa}(\tau)$ encodes information about the equilibrium
dynamics of the system through 
its connection to the dynamic local 
susceptibility $\chi(\omega)$,
\begin{equation}
\label{connection}
Q_{aa}(\tau) = \int_{-\infty}^{\infty} \frac{d\omega}{\pi} \chi''(\omega)
\frac{
\exp(-\omega |\tau|)}{1 - \exp(-\beta \omega)}
\; ,
\end{equation}
where  $\chi''(\omega)=\mbox{Im}\chi(\omega)$. 
It follows that $\tilde{Q}_{aa}(i \omega_k) = 
\left. \chi(\omega)\right|_{\omega +
i \epsilon \to i \omega_k}$ and $Q_{aa} = 
\left. C(t)\right|_{t \to -
i \tau}$, where $C$ is the 
real-time autocorrelation function \cite{footnote}.
We shall hereafter 
use the notation $Q_{aa}(\tau)=q_d(\tau)$.

The off-diagonal elements $Q_{ab}$ are $\tau$-independent.
 This is  
a general and important property of quantum disordered 
systems in equilibrium proven by Bray and Moore 
in their pioneering work on quantum spin-glasses
\cite{bm-quantum}. 
The argument goes as follows. The overlap matrix for $a\neq b$ 
is
\begin{equation}
N Q_{a\neq b} (\tau,\tau') = 
\overline{\langle \vec s_a(\tau) \cdot \vec s_b(\tau') \rangle }
\; .
\end{equation}
Before performing the average over disorder, replicas are decoupled and 
the thermal average factorizes. But the averages $\langle
s_a(\tau)\rangle$  are time-independent for any fixed configuration of
disorder and so is the average of their product.  
 Consequently, the 
Fourier transform of $Q_{a\ne b}(\tau)=q_{a\ne b}$ 
is given by
\begin{equation}
{\tilde{q}}_{a\ne b} = q_{a\ne b}\hbar\beta\delta_{\omega_k,0}.
\label{eq:offd}
\end{equation}
We conclude that the {\sc rsb} 
solution is confined to the zero mode of the 
Matsubara frequencies and that $q_{a\ne b}$ is the quantum analog of
the classical order-parameter matrix.

\subsection{Relationships to other models}
\label{comparison}

Besides being a quantum extension of the spherical $p$
spin-glass this model can be interpreted in other ways. First,
if we identify $\vec{s}=(s_1,s_2,\ldots,s_N)$ as a position vector in
$N$-dimensional space, Hamiltonian (\ref{eq:action}) with the
constraint (\ref{eq:lm}) describes the 
motion of a quantum particle 
of mass $M$ constrained to move on the $N$-hypersphere of radius 
$\sqrt{N}$ in the presence of a
random potential $V(\vec{s})$ 
with $p$-dependent correlations given by Eq.~(\ref{corr_pot}).

Second, the well known connection between quantum mechanics in $D$ dimensions 
and statistical mechanics in $D+1$ dimensions allows us to interpret 
this model as representing a classical closed polymer in a
quenched random medium in infinite transverse 
dimensions~\cite{Gold3,Ha-he,Baum}.
The imaginary time $\tau$ represents the 
internal coordinate of the polymer of contour length $L=\beta\hbar$. 
The low temperature limit of the quantum model corresponds then to 
the long length limit of the polymer.
The $N$ components of the 
field $\vec s$ denote the $N$ transverse coordinates on the 
spherical embedding 
space. The parameter $\Gamma$ is a measure of the linear 
elasticity of the polymer since the first term on the left-hand-side of
Eq.~(\ref{eq:motion2}) may be thought of as 
deriving from an elastic energy. 
Finally, the last interaction term is a consequence of the 
average over the quenched disorder. The time-dependent 
replica matrix $Q_{ab}(\tau)$ quantifies the correlation between the 
position of different monomers, $\tau$ being the distance between
them, measured following the polymer direction. 
The replica indices represent different states of the 
polymer. 

There are also several quantum models related to this one.
When $p=2$ one recovers the model studied in Ref.~\cite{p2} as well as 
the disordered quantum-rotor model of Ref.~\cite{Sachdev}. 
Quantum Ising spin models with $p$-spin interactions 
in a transverse field have 
been extensively studied in the literature 
\cite{Gold,Niri,p-spin-transv}.  
A different quantization rule for spherical spins was proposed in 
Ref.~\cite{Ni} and it was later studied in detail, including $p$-spin interactions, in Ref.~\cite{Niri}.
The difference in the quantization rule results in a different
imaginary-time dynamics as the equation of motion contains first-order 
imaginary-time derivatives instead of second-order
ones as we have here.

Notice that if we set $p=4$, Eq.~(\ref{eq:motion2}) becomes similar to  
Eqs.~(4)-(6) in Ref.~\cite{antoine} 
for the replicated boson Green's function $G_{ab}(\tau)$ 
in the SU(${\cal N}$) Heisenberg spin-glass model. 
Indeed, the equations 
look identical if we formally take $S=-1$, redefine $J$ and  replace the
term linear in 
the Matsubara frequencies by a quadratic one, {\it i.e.}, if we use  
first-order instead of second-order imaginary-time derivatives. 
Note, however, that the symmetry properties of the quantities $G$ (in
Ref.~\cite{antoine})
and $Q$ (here) are quite different. 

Chandra {\it et al.} showed the equivalence between a model for a
 classical array of  Josephson junctions in a magnetic field 
and the $p$-spin-glass model 
with $p=4$~\cite{Premietal}. Using 
a diagrammatic expansion technique similar to that 
used to derive classical or quantum TAP equations,  
Kagan {\it et al.}~\cite{Kagan}
showed that the equations governing the equilibrium behavior 
of the quantized model  
are similar, though not identical, to 
the Matsubara equations  we analyze here. 
  
\section{The static approximation}
\setcounter{equation}{0}
\renewcommand{\theequation}{\thesection.\arabic{equation}}
\label{static_approx}

Before presenting the full numerical solution of the model it is
useful to study it using a very simple approximation known as the 
static approximation. 
The latter, first introduced as a variational
{\it Ansatz} in
Ref.~\cite{bm-quantum}, consists in proposing $q_d(\tau)\equiv
q_d$, independent of $\tau$, and determining $q_d$ by minimization of the free-energy.
Despite its simplicity, detailed comparison with exact numerical
results \cite{large_D,Niri} has shown that the static approximation 
captures much of the physics of quantum spin mean-field models.
In our case, however, the na\"{\i}ve static approximation applied at the level
of the saddle point equations is oversimplified 
since  the spherical constraint
(\ref{eq:sc}) immediately sets $q_d=1$ thus loosing all variational freedom.
We then proceed differently and replace $q_d(\tau)$ by $q_d$ and
$\lambda_d(\tau)$ by $\lambda_d$ at the level of the effective action 
(\ref{eq:seff}) and 
 treat both quantities as variational parameters. 
In this way, we do not impose the spherical constraint strictly
(since $q_d$ can be different from $1$) but in an averaged manner. 

The effective action in the static
approximation becomes
\begin{eqnarray}
-\frac{1}{\hbar} S_{\sc eff} &=&
\frac{N}{2} \mbox{Tr} \ln {\bf Q} - 
\frac{Nn}{2} \sum_{k\neq 0} 
\ln\left[ \beta \left( \frac{\omega^2_k}{\Gamma} +z \right) \right]
+ \frac{N \beta^2}{4} \sum_{ab} Q_{ab}^{\bullet p}
\nonumber\\
& & 
+\frac{Nn}{2} \left( 1 +\beta z (1-q_d)\right) 
\; .
\end{eqnarray}
By using the identity~\cite{Fe}
\begin{equation}
\sum_{k} \ln \left[ \beta \left( \frac{ \omega_k^2}{\Gamma} + z
\right) \right]
= 
2 \ln \left[ 2 \sinh \left( \frac{\beta\sqrt{\Gamma z}}{2} \right)\right]
\; , 
\end{equation}
we get the following expression for the free-energy density
\begin{eqnarray}
\beta f &=& 
-\frac{1}{2}
\ln(\beta z) + 
\ln \left[ 2 \sinh \left( \frac{\beta
\sqrt{\Gamma z}}{2} \right)\right] - 
\lim_{n\to 0}
\frac{1}{n} \left[ 
 \frac{1}{2} \mbox{Tr} \ln {\bf Q} 
+
\frac{\beta^2}{4} \sum_{ab} Q_{ab}^{\bullet p}
\right]
\nonumber\\
& & - \frac{1}{2} \left( 1 +\beta z (1-q_d)\right) 
\; .
\label{eq:free_static}
\end{eqnarray}
The saddle-point equation with respect to $z$ reads
\begin{equation}
q_d = 1 + \frac{1}{\beta z} - \frac12 \sqrt{\frac{\Gamma}{z}}
\coth\left( \frac{\beta}{2} \sqrt{\Gamma z} \right)
\label{eq:static_ap_z}
\end{equation}
and replaces the spherical constraint. As 
expected, it is independent of the non-diagonal part
of the ${\bf Q}$ matrix.
At high temperature ($\beta \to 0$) one recovers the classical result,
$q_d\equiv 1$. In the opposite limit, $\beta \to \infty$, 
\begin{equation}
q_d \to 1 - \frac{1}{2} \sqrt{\frac{\Gamma}{z}}
\; .
\label{eq:zero-point}
\end{equation}
The second term on the right hand side of the equation corresponds to
 the
zero-point reduction of the order parameter characteristic of  
quantum magnets.
 
The saddle-point equation with respect to the order-parameter 
$Q_{ab}$ is
\begin{equation}
z\delta_{ab} = \frac{1}{\beta} Q^{-1}_{ab} 
+ 
\frac{p\beta}{2} Q_{ab}^{\bullet \,p-1}
\; .
\label{eq:static_sa_qab}
\end{equation} 

\subsection{The paramagnetic solution}

At high temperatures or if quantum fluctuations are strong 
we expect a {\sc pm} phase.  
Within the static approximation, the {\sc pm} solution is 
associated to a diagonal replica matrix ${\bf Q}$:
\begin{equation}
Q_{ab}(\tau) = q_d \delta_{ab}
\; .
\end{equation}
The saddle-point equation (\ref{eq:static_sa_qab}) is now very simple
\begin{eqnarray}
z &=& \frac1{\beta q_d} + \frac{p \beta}{2} q_d^{p-1}
\; .
\label{eq:z_static_pm}
\end{eqnarray}
This equation and Eq.~(\ref{eq:static_ap_z}) form a set of two  
coupled equations for $q_d$ and $z$ that can be solved
numerically for all $\beta$ and $\Gamma$.

Some asymptotic behaviors can be obtained analytically.  
First, in the classical limit, $T$ fixed and $\Gamma\to 0$, 
Eqs.~(\ref{eq:static_sa_qab}) and (\ref{eq:z_static_pm}) imply 
$q_d=1$ and $\beta z = 1+p\beta^2/2$, in agreement with the results in  
Ref.~\cite{crisanti}.
Second, when $T \gg 1$ but $\beta \Gamma \ll 1$, 
quantum fluctuations are still irrelevant and we get $\beta z = 1$ and $q_d=1$.
Third, in the limit $\Gamma\to\infty$ and $T$ finite, $q_d$ tends to zero 
as $q_d\sim 4/(\beta\Gamma)$ and $z\sim \Gamma/4$. 
The zero-temperature limit is more subtle and we discuss it below. 

\begin{figure}
 \centerline{\hbox
{   \epsfig{figure=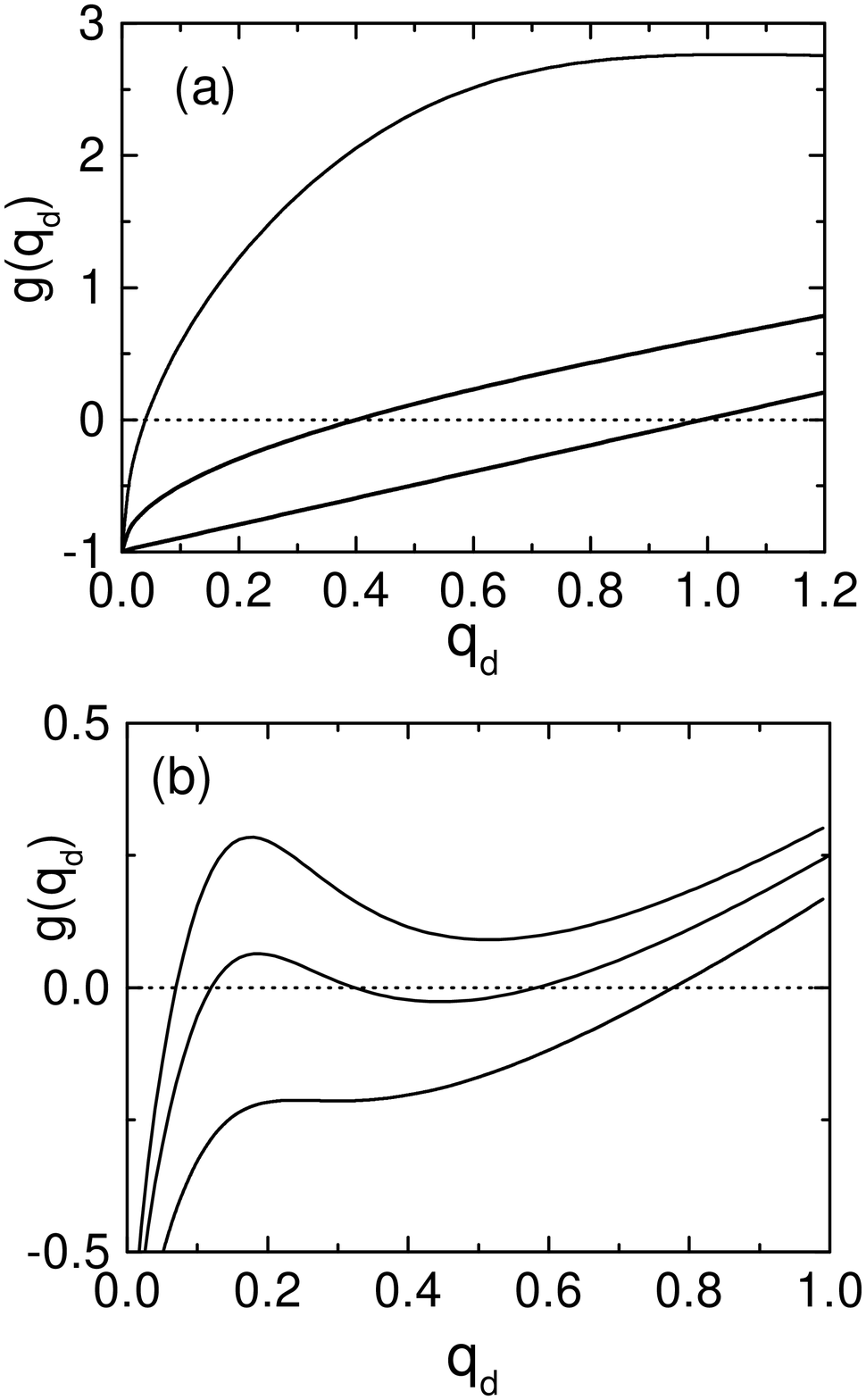,width=8cm}}
 }
 \caption{(a): The function $g(q_d)$ that determines the {\sc pm} solutions 
for $\beta~=~1<~\beta_{\sc p}\approx~6$ and $\Gamma = 0.1,10,100$, from
bottom to top.
For all $\Gamma$ we find only one solution that moves towards 
smaller values of $q_d$ when $\Gamma$
increases. (b): The same for 
$\beta=10>\beta_{\sc p} \approx 6$ and several 
choices of $\Gamma$. For $\Gamma=6>\Gamma_{c1}$ (top curve) there is
only one solution with a small value of $q_d$; for 
$\Gamma_{c2} < \Gamma=4 < \Gamma_{c1}$ (middle curve) there are three solutions, while
for $\Gamma=2 < \Gamma_{c2}$ (lower curve) there is again only one solution but with a large
value of $q_d$.}
 \label{fig:pm_static}
\end{figure}

The properties of the general solution of
Eqs.~(\ref{eq:static_sa_qab}) 
and (\ref{eq:z_static_pm}) are best understood 
graphically. 
In Figs.~\ref{fig:pm_static}(a) and (b)  we display 
$g(q_d) \equiv q_d - $ rhs(Eq.~(\ref{eq:static_ap_z})) {\it vs.} $q_d$, 
obtained replacing 
$z$ by its value given by Eq.~(\ref{eq:z_static_pm}), for several 
values of $T$ and $\Gamma$. The points
where $g(q_d)$ crosses the horizontal axis are the solutions 
for $q_d$. 

The summary of our results for $p\geq 3$ is the following 
(the case $p=2$ is different and we discuss it in Section~\ref{p2_sect}):

\begin{itemize}

\item 
For fixed $\beta<\beta_{\sc p}$ ($\approx 6$ for $p$ = 3) 
and all values of  $\Gamma$, the equations
 admit only
one solution with $0\leq q_d \leq 1$. The function $q_d(\Gamma)$ 
 decreases monotonically with $q_d(0) = 1$ and 
$q_d\to 0$ when  $\Gamma\to \infty$. This is represented in 
Fig.~\ref{fig:pm_static} (a), where we show
$g(q_d)$ for $\beta=1<\beta_{\sc p}$ and  
$\Gamma=0.1,10,100$. The analysis of the 
free-energy density shows that this solution is a local minimum of $f(q_d)$.
 
\item
For $\beta>\beta_{\sc p}$, there are several {\sc pm}   
solutions and  phase transitions can occur between them upon varying $\Gamma$. 
For $\Gamma$ above a critical value $\Gamma_{c1}$, there exists a 
solution {\sc pm}$_1$ with a small value of $q_d = q_d^<$  
(the upper curve in Fig.~\ref{fig:pm_static}(b)) that we  call a 
quantum paramagnet.
 This solution can be continued below $\Gamma_{c1}$ but two new ones
appear: {\sc pm}$_2$, that we call a classical paramagnet, with large
$q_d = q_d^> $, 
and  {\sc pm}$_3$ with an  
intermediate value of $q_d $. 
The three solutions coexist in the interval $\Gamma_{c2} <\Gamma < \Gamma_{c1}$.
At $\Gamma_{c2}$,  {\sc pm}$_1$ and  {\sc pm}$_3$ 
merge and disappear. The three coexisting 
solutions are shown in the intermediate
curve in Fig.~\ref{fig:pm_static}(b). Below $\Gamma_{c2}$, only {\sc
pm}$_2$ exists (the lower curve in Fig.~\ref{fig:pm_static}(b)). 
An analysis of the stability of these solutions shows that, where they
exist, 
 {\sc pm}$_1$ and {\sc pm}$_2$ are local minima of the 
free-energy whereas  {\sc pm}$_3$ is a local maximum and can 
therefore be immediately discarded.
In the limit $\beta\to\infty$
the two stable solutions can be simply computed and 
one obtains $q^>_d\sim 1-\sqrt{\Gamma/(2 p)} \beta^{-1/2} \to 1$ with 
$z\sim p\beta/2$ and  
$q^<_d \sim 4/(\Gamma\beta) \to 0$
 with 
$z\sim \Gamma/4$. Both solutions exist for all 
values of $\Gamma$ when $T\to 0$, indicating that 
$\Gamma_{c1} \to \infty$ and $\Gamma_{c2} \to 0$ as $T\to 0$.
The dotted lines in Fig.~\ref{fig:phasedia} delimit the region where there is 
coexistence of paramagnetic solutions in the case $p=3$.

\item
The free-energies of the two stable {\sc pm} solutions cross 
at $\Gamma_{c0}$, a coupling intermediate 
between $\Gamma_{c1}$ and $\Gamma_{c2}$. 
For  $\Gamma > \Gamma_{c0}$, $f(q^<_d) < f(q^>_d)$ but the 
inequality is reversed  for $\Gamma < \Gamma_{c0}$. 
This would suggest that, for all $0 \le T \le T_{\sc p}$, 
there is a first-order transition between the 
 two stable {\sc pm} states with  {\sc pm}$_2$ favored 
for $\Gamma < \Gamma_{c0}$. However, we will see below
 that this transition is only possible for $T^{\star}\le
T \le T_{\sc p}$ where $T^{\star}$ is slightly below $T_{\sc p}$ for small to
moderate $p$.
\end{itemize}

\subsection{The {\sc sg} solution}
\subsubsection{The replica symmetric solution}
At low temperatures one expects the system to exhibit 
a non trivial {\sc sg} solution characterized by a replica matrix with 
non-vanishing off-diagonal elements. 
The simplest possible {\it Ansatz} is the replica-symmetric ({\sc rs}) one,
\begin{equation}
Q_{ab} = (q_d-q_{\sc ea}) \delta_{ab} + q_{\sc ea} .  
\end{equation} 

From the general expression~(\ref{eq:free_static}) for the 
free-energy one then 
derives three equations for $z,q_d$ and $q_{\sc ea}$. The 
equation for $z$ is independent of the {\it Ansatz} and hence coincides 
with Eq.~(\ref{eq:static_ap_z}). The other two equations read
\begin{eqnarray}
z &=& \frac{1}{\beta} \frac{q_d-2 q_{\sc ea}}{(q_d-q_{\sc ea})^2} + 
\frac{\beta p}{2} q_d^{p-1}\; ,
\label{eq:qd_static_rs}
\\
0 &=& -\frac{1}{(q_d - q_{\sc ea})^2} + 
\frac{p \beta^2}{2} q_{\sc ea}^{p-2}
\; ,
\label{eq:q1_static_rs} 
\end{eqnarray}
after we eliminated  the paramagnetic solution  $q_{\sc ea}=0$. 

These equations have non-trivial solutions in a finite region 
of phase space satisfying $0\leq q_{\sc ea} \leq q_d$. The phenomenon 
of multiplicity of solutions at low temperatures found in
the {\sc pm} case appears also here. 
For temperatures that are not too low, ({\it e.g.} for 
$\beta=10$ for $p=3$), these equations have two solutions, one 
stable and  one unstable {\it within} the {\sc rs} {\it Ansatz}. At very low 
temperatures, two other solutions appear, one stable and 
one unstable within {\sc rs}. 
However, a careful analysis of stability shows that, as in the 
classical case \cite{crisanti}, the {\sc rs} solutions are always {\it
unstable} with respect 
to replica symmetry breaking. To show this, it is sufficient to 
compute the {\it replicon} $\Lambda_T$, the transverse
eigenvalue of the Hessian matrix . We will derive a general expression for
$\Lambda_T$ in 
Section~\ref{derivation_replicon}. Here we simply quote the result in the 
static approximation which is

\begin{equation}
\Lambda_T =\frac{1}{(q_d-q_{\sc ea})^2}-
{\beta^2 \over 2}p(p-1)q^{p-2}_{\sc ea}.
\label{eq:repstat} 
\end{equation}
It follows from this equation and Eq.~(\ref{eq:q1_static_rs})
that $\Lambda_T$ cannot be positive for any of the {\sc rs} solutions
if $p\geq 3$. (See Section~\ref{p2_sect} for a discussion of the case $p=2$.)

\subsubsection{The 1-step replica symmetry breaking solution}
\label{1steprsbsol}

Using Parisi's {\sc rsb} scheme, 
we search now for a 1-step ({\sc rsb}) 
solutions of the form
\begin{equation}
Q_{ab} = (q_d-q_{\sc ea}) \delta_{ab} + (q_{\sc ea} -
q_0) \epsilon_{ab} + q_0\; ,
\end{equation} 
where the matrix ${\bf \epsilon}$ is defined as
\begin{equation}
\epsilon_{ab}=\left\{
\begin{array}{l}
1 \;\;\;\;\; \mbox{if} \; a \; \mbox{and} \; b \; \mbox{are in a diagonal block} \\
0 \;\;\;\;\; \mbox{otherwise}
\end{array}
\right.
\label{eq:eps_ab}
\end{equation}
We show in the Appendix that this {\it Ansatz} actually 
gives the exact solution to the full problem. 
In the absence of an external magnetic field $q_0=0$ and 
the extremization equations with respect to $q_d$, $q_{\sc ea}$ and $m$
read
\begin{eqnarray}
\beta z &=&  \frac{q_d+(m-2)q_{\sc ea}}{(q_d-q_{\sc ea}(1-m))(q_d-q_{\sc ea})}
+ \frac{p\beta^2}{2} q_d^{p-1}
\; , 
\label{eq:qdeq_static_sg}
\\
0 &=& \frac{1}{(q_d-q_{\sc ea}(1-m))(q_d-q_{\sc ea})} - \frac{p \beta^2}{2} q_{\sc ea}^{p-2} 
\; , 
\label{eq:q1eq_static_sg}
\\
0 &=& \frac{p q_{\sc ea}}{m(q_d-q_{\sc ea}(1-m))} + \frac{p}{m^2} 
\ln \left(\frac{q_d-q_{\sc ea}}{q_d-q_{\sc ea}(1-m)}\right) + \frac{p\beta^2}{2}
q_{\sc ea}^p
\; , 
\label{eq:meq_static_sg}
\end{eqnarray}
where we excluded the two solutions $m=1$ or $q_{\sc ea}=0$ from 
Eq.~(\ref{eq:q1eq_static_sg}). 
The first and second equations above have new
$m$-dependent factors. These equations 
reduce to the {\sc rs}
equations when  
$m=0$. The third equation is new and is used to determine 
the break point $m$.  

It is useful to define the parameters
\begin{equation}
y=\frac{q_{\sc ea}}{q_d} \; , \;\;\;\;\;\;\;\; x_p = \frac{my}{1-y}
\; ,
\label{eq:2par}
\end{equation} 
Subtracting 
Eq.~(\ref{eq:meq_static_sg}) from Eq.~(\ref{eq:q1eq_static_sg}) 
and writing the result in terms of $x_p$ one obtains
\begin{equation}
\ln\left(\frac{1}{1+x_p}\right) + \frac{x_p}{1+x_p} + \frac{x_p^2}{p(1+x_p)} =0
\; .
\label{eq:xxx}
\end{equation}
This is a master equation for $x_p$ that 
only depends on the parameter $p$. 
Once $x_p$ is known from this equation, $y$ follows from $y=x_p/(m+x_p)$.
For $p=3$, a case that we will consider in detail
\begin{equation}
x_3=1.81696
\end{equation}

Equation~(\ref{eq:q1eq_static_sg}) yields a simple relation
between $q_{\sc ea}$ and $x_p$. Using then $q_{\sc ea}=x_p/(m+x_p) q_d$
one has 
\begin{eqnarray}
\frac{p(m\beta)^2}{2}  q_{\sc ea}^p &=& \frac{x_p^2}{1+x_p}
\; ,
\label{eq:q1_x_stat}
\\
\frac{p(m\beta)^2}{2}  q_d^p &=& \frac{x_p^{2-p} (m+x_p)^p}{1+x_p} 
\; .
\label{eq:qd_x_stat}
\end{eqnarray}

For fixed values of $m$ and $\beta$, one obtains $q_{\sc ea}$, $q_d$ and $z$ 
from Eqs.~(\ref{eq:q1_x_stat}), (\ref{eq:qd_x_stat}) and (\ref{eq:qdeq_static_sg}), 
respectively. Solving numerically Eq.~(\ref{eq:static_ap_z}) one gets 
the corresponding value of $\Gamma$. 

To study the transitions between the  {\sc sg} and
{\sc pm} phases described above we compare their free energies. 
 Three cases
must be distinguished

\begin{figure}
 \centerline{\hbox
{   \epsfig{figure=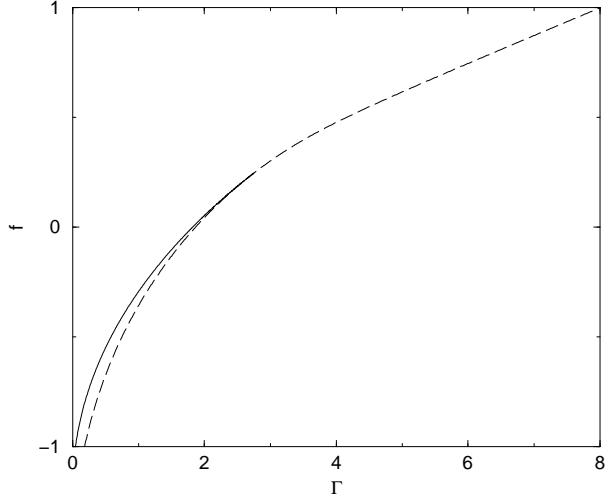,width=8cm}}
 }
\caption{The {\sc sg} and {\sc pm} free-energies (solid and dashed lines, respectively) 
in the static approximation as a function of $\Gamma$ for $\beta=4$ and $p=3$.}
 \label{fig:SGPM4}
\end{figure}

\begin{figure}
 \centerline{\hbox
{   \epsfig{figure=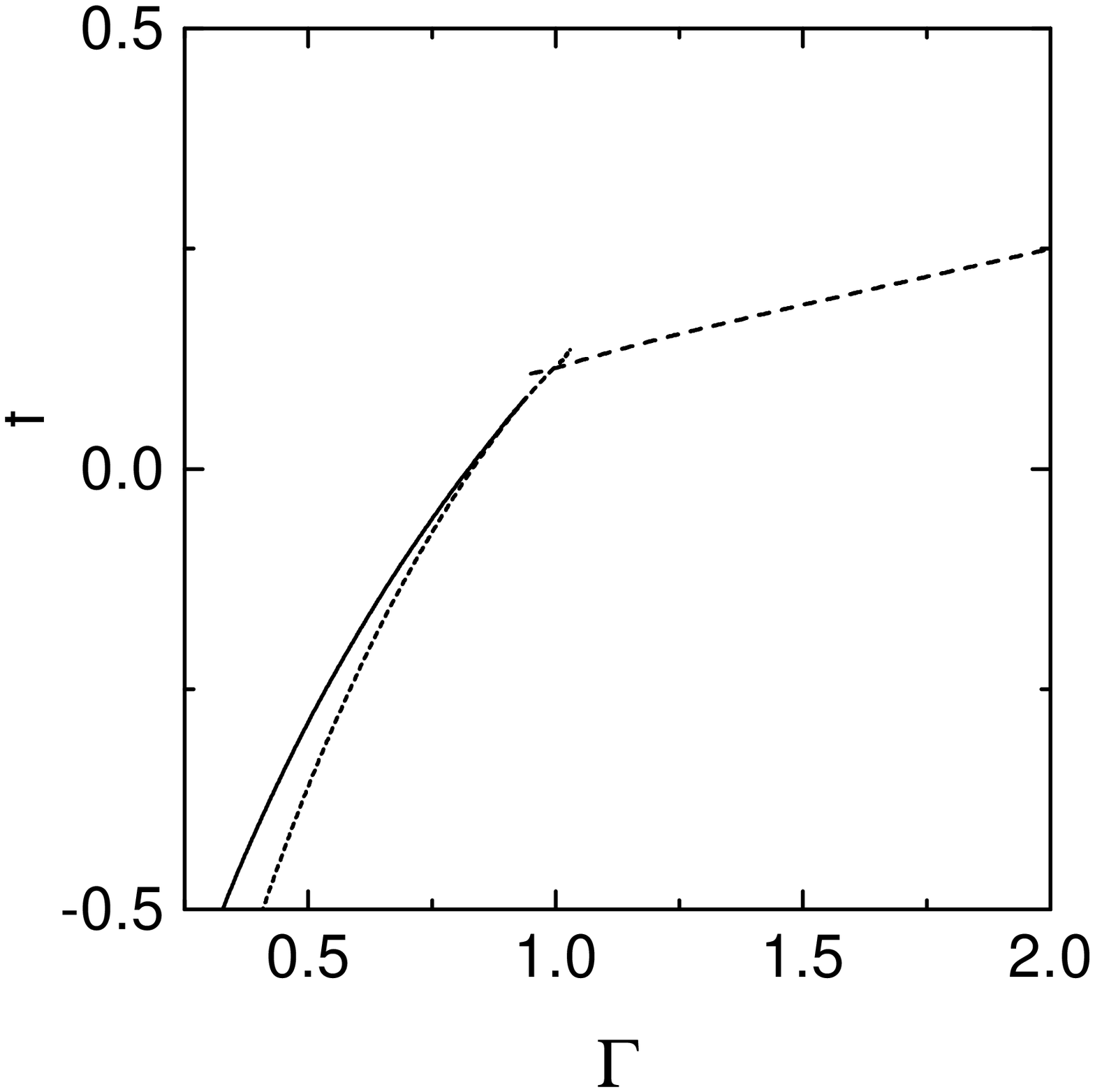,width=8cm}}
 }
 \caption{The {\sc sg} free-energy (solid line) and the free-energies of 
{\sc pm}$_1$ and 
{\sc pm}$_2$ 
(long-dashed and dashed lines, respectively) in the static approximation 
as a function of $\Gamma$ for $\beta=6.5$ and $p=10$. In this case $f_{\sc sg}$
intersects $f_{{\sc pm}_2}$.}
 \label{fig:SGPMp10B65}
\end{figure}

\begin{figure}
 \centerline{\hbox
{   \epsfig{figure=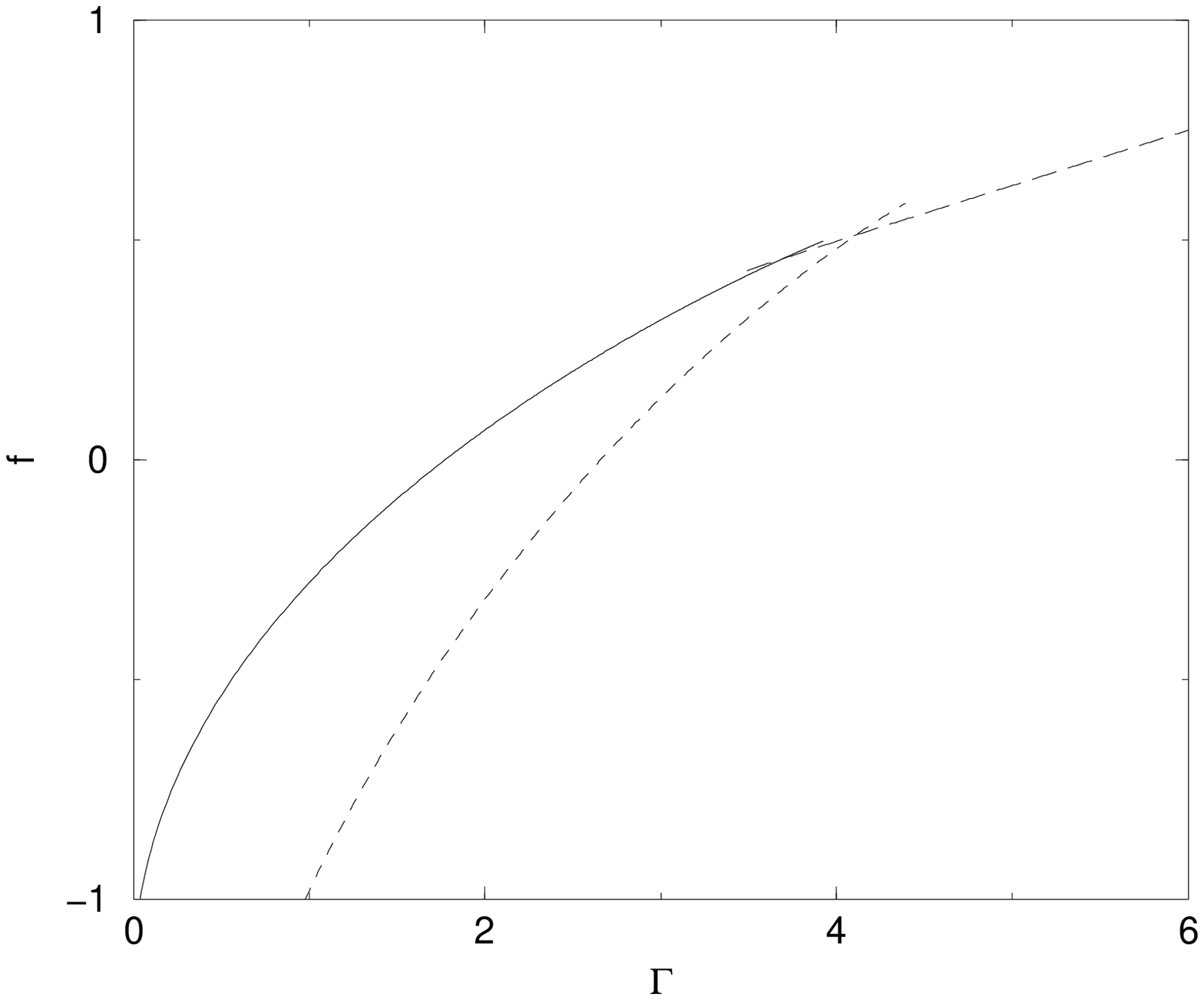,width=8cm}}
 }
 \caption{The {\sc sg} free-energy (solid line) and the free-energies of 
{\sc pm}$_1$ and 
{\sc pm}$_2$ 
(long-dashed and dashed lines, respectively) in the static approximation 
as a function of $\Gamma$ for $\beta=10$ and $p=3$.  In this case $f_{\sc sg}$
intersects $f_{{\sc pm}_1}$.}
 \label{fig:SGPM10}
\end{figure}

\begin{itemize}

\item 
$\beta < \beta_{\sc p}$ (but greater than the inverse of the classical 
critical temperature). This case is represented in
Fig.~\ref{fig:SGPM4} for $p=3$ and $\beta = 4$. The {\sc sg} and {\sc
pm} free energies increase monotonically with $\Gamma$. At $\Gamma
=\Gamma_c$, $f_{\sc sg}=f_{\sc pm}$ and a phase transition  occurs. 

As in the classical case (where the temperature is the control
parameter), $m=1$ at the transition. The spin-glass solution continues to exist beyond 
the transition point until  $\Gamma = \Gamma_{\sc max} > \Gamma_c$ 
where it disappears. However, between $\Gamma_c$ and  $\Gamma_{\sc max}$, 
 $m > 1$ . Values of $m$ greater than 1 do not represent
physically allowed states and should not be considered.  
This follows from the fact that the susceptibility 
$\chi = \beta \left[q_d - (1 -m) q_{\sc ea}\right]$ can be shown to 
be strictly {\it smaller} than  $\beta q_d$, which implies $0 \le m
\le 1$. 
Below the critical point,
$f_{\sc sg} >f_{{\sc pm}}$ meaning that the {\sc sg} solution
{\it maximizes} the free-energy. 

The {\sc pm} solution exists below the transition  
all the way down to $\Gamma=0$. This solution is locally stable  
for all values of $\Gamma$ and could be interpreted as
 a metastable state below $\Gamma_c$. 
We believe, however, that this solution should be discarded on
physical grounds because it is continuosuly connected to a $\Gamma =0$
state whose 
 thermodynamic properties at low temperatures 
are unphysical: the internal energy and the 
susceptibility of this state diverge as $T \to 0$ whereas these
quantities  can be shown to be finite in the ground state of
Hamiltonian~(\ref{eq:action}). Therefore, we shall consider that  
 the continuation of the {\sc pm} solution below
$\Gamma_c$ is a 
 spurious solution and that the only physically allowed phase
below $\Gamma_c$ in this temperature range is the {\sc sg}.

The Edwards-Anderson order parameter $q_{\sc ea}$ is
{\it discontinuous} at the transition.
 The transition is nevertheless 
 of {\it second}
order in the thermodynamic sense. This follows from the fact that 
$m=1$ at $\Gamma_{c}$ and so the effective number of
degrees of freedom involved is $(1-m) q_{\sc ea} \to 0$ 
at $\Gamma_{c}$. Therefore, there is no latent heat 
and the linear susceptibility  
is continuous. However, in contrast with the SK-model, 
the transition is not associated with a divergence of the 
{\sc sg} susceptibility. 
 The situation is the same as in the replica theory of the 
classical model \cite{crisanti}.  

\item
$\beta^{\star}< \beta <\beta_{\sc p}$ where 
$T_{\sc p} - T^{\star} \ll T_{\sc p}$
except for very large values of $p$. The case $p=10$  for $\beta = 6.5$ 
is represented in
Fig.~\ref{fig:SGPMp10B65} where we show
the free-energies of the two {\sc pm} and  the {\sc sg} solutions as
functions of $\Gamma$. Decreasing $\Gamma$ from $\Gamma \gg 1$
we encounter two phase transitions. The first one occurs when the free 
energies of the two
{\sc pm} phases cross which signals a first order phase transition
between the quantum and classical paramagnets. 
The glass transition occurs at a lower coupling, 
when the free energies of the {\sc pm}$_2$ and {\sc sg} states
cross. This transition is similar to that found for 
$T \ge T_{\sc p}$ and it is characterized by a discontinuity of the order
parameter and the value $m=1$ at the transition. As in the previous case, 
$f_{\sc sg} >f_{{\sc pm}_2}$ and the {\sc sg} solution
{\it maximizes} the free-energy. 

\item
$\beta > \beta^{\star}$. This case is  represented in Fig.~\ref{fig:SGPM10}
for $p=3$ and $\beta=10$. The  free-energies of the
states {\sc pm}$_1$ and  
{\sc pm}$_2$ cross as before. However, $f_{{\sc sg}}$ does not cross $f_{{\sc
pm}_2}$ but it crosses $f_{{\sc pm}_1}$ instead. 
Therefore, a transition to the low-temperature ordered state can occur
from {\sc pm}$_1$,  not from {\sc pm}$_2$.  
If we were to interpret the crossings of the free energies in the standard
way, we would conclude that below the
transition to {\sc pm}$_2$ the system stays in that state 
down to $\Gamma =0$. However,
{\sc pm}$_2$, which 
is the continuation to low temperatures of the {\sc pm}
state discussed above for $\Gamma < \Gamma_c$ and $T > T_{\sc p}$, 
exhibits the same unphysical
properties. We thus conclude that, for $\beta>\beta^{\star}$, the {\sc pm}$_1$ to 
{\sc pm}$_2$ transition is spurious and that the system stays in the {\sc
pm}$_1$ state until the latter disappears in favor of the  {\sc sg}
state. It should be noted that a similar situation was encountered
in a study of the Ising {\sc sg} model in a transverse field at 
low temperatures \cite{Rogr}. In contrast to the situation for $\beta
< \beta^{\star}$, $f_{\sc sg} < f_{{\sc pm}_1}$ and the {\sc sg} solution
{\it minimizes} the free-energy. 
Such an inversion of the order of the free-energies was also
 found in a study of an anisotropic classical $p$-spin model 
\cite{mottishaw}. It can be thought as a result of the competition between 
the conflicting requirement of maximization and minimization with 
respect to $q_{\sc ea}$ and 
$q_d$, respectively. 

We find that the break point $m$  considered 
as a function of $\Gamma$ has  two branches. 
Physical values of $m$ must be chosen from the branch that 
can be continued to the classical limit, $\Gamma=0$.
For $\beta > \beta^{\star}$, the 
maximum physical value of the break point $m_{\sc max} < 1$. 
The {\sc sg} solution 
ceases to exist at the corresponding value of $\Gamma$, $\Gamma_{\sc max}$ 
(this will become more transparent in the zero temperature limit 
that we discuss next). An important consequence is that now $m$ is 
discontinuous at the
transition implying that the latter becomes {\it
first} order. Since the physical {\sc sg} and {\sc pm}
solutions extend beyond the point where their free-energies cross, 
there is a region of
phase coexistence and hysteresis is expected (see below).
\end{itemize}

\begin{figure}
 \centerline{\hbox
{   \epsfig{figure=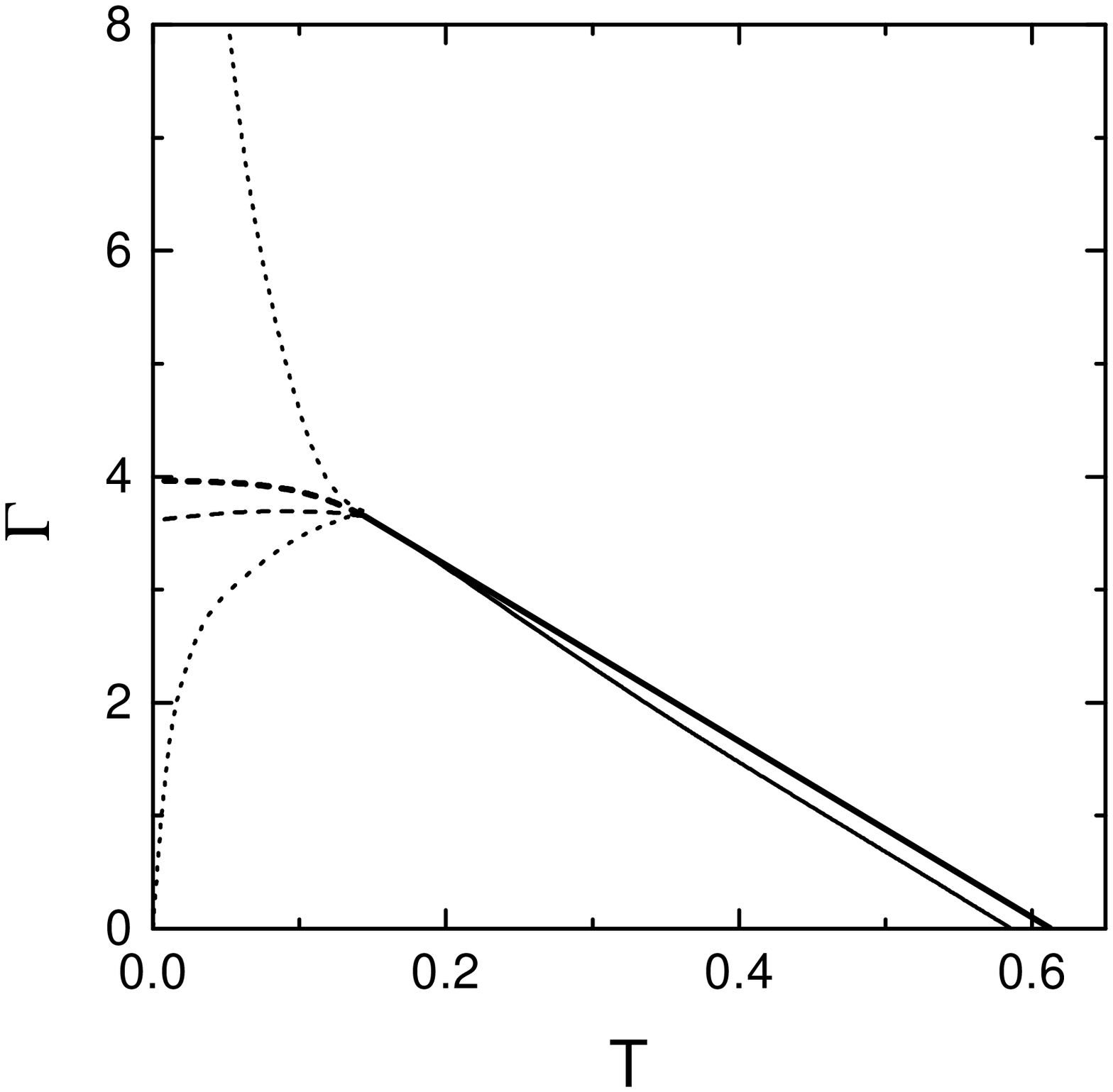,width=8cm}}
 }
 \caption{Static (thin lines) and dynamic (thick lines) phase diagrams 
of the $p$-spin model for $p=3$ in the static approximation. 
Solid and dashed lines represent second 
and first order transitions, respectively. The dotted line delimits
the region where two stable paramagnetic solutions coexist.}
 \label{fig:phasedia}
\end{figure}

The phase diagram resulting from this analysis is
represented for $p=3$ in Fig.\,\ref{fig:phasedia} (thin lines). The
{\sc pm}$_1$ to {\sc pm}$_2$ transition is invisible on the scale  of the plot
for this value of $p$. 
At low temperatures, the first order line exhibits reentrant behavior, 
as is the case for the Ising $p$-spin-glass model in a transverse field \cite{Niri}. 
This feature is an artifact of the static approximation that
disappears in the exact treatment of the problem, as we shall see 
in  Section~\ref{exact}. 
The second and first order critical lines meet at 
the tricritical point $(T^{\star},\Gamma^{\star})$.
It is interesting to notice that the $180^o$ rule \cite{Wheeler} 
that imposes that the angle 
formed by the intersection of two critical lines at a tricritical point 
should be $<180^o$ is satisfied in the static approximation. 

\subsubsection{The static approximation in the zero temperature limit}

The behavior of the system can be completely elucidated at 
zero temperature. In this limit, the requirement that $q_{\sc ea}$ is
finite implies
\begin{equation}
\beta m  < +\infty
\; ,
\end{equation} 
meaning that $m\to 0$. This implies that 
Eq.~(\ref{eq:qd_x_stat}) can be simplified to 
\begin{eqnarray}
\frac{p}{2} (\beta m)^2 q_d^p &\sim& \frac{x_p^2}{1+x_p}
\end{eqnarray}
and $\coth(\beta\sqrt{\Gamma z}/2)$ in Eq.~(\ref{eq:static_ap_z})
can be replaced by one. The Lagrange multiplier in 
Eq.~(\ref{eq:qdeq_static_sg}) 
becomes
\begin{equation}
z \sim \frac{x_p^2}{(1+x_p) q_d \beta m}
\label{eq:z_eq_T0}
\end{equation}
and we obtain for $\Gamma$ as a function of $q_d$,
\begin{equation}
\Gamma = \sqrt{\frac{8 p x_p^2}{1+x_p}}
q_d^{(p-2)/2} \left( 1 - q_d\right)^2
\; .
\label{eq:Gamma_qd_eq_T0}
\end{equation}
The right hand side (rhs) of this equation is a bell-shaped curve,
that vanishes at 
$q_d=0$ and $q_d=1$. It reaches its maximum at $q_d^{\sc max}=(p-2)/(p+2)$,
independently
of $x_p$. The corresponding value of $\Gamma$ is  

\begin{equation}
\Gamma^{\sc static}_{\sc max} = \sqrt{\frac{8 p x_p^2}{1+x_p}}
\left(\frac{p-2}{p+2}\right)^{(p-2)/2} \left(\frac{4}{p+2}\right)^2
\; . 
\end{equation}
The physically meaningful solution for 
$\Gamma < \Gamma^{\sc static}_{\sc max}$ must be searched  
 on the branch  
corresponding to $q_d>q^{\sc max}_d$. This is the branch that gives
the correct classical limit,  $q_d = 1$ for $\Gamma=T=0$ and has the expected property that $q_{\sc ea}$ is a decreasing function of $\Gamma$.
This solution can be continued until $\Gamma = \Gamma^{\sc
static}_{\sc max}$. However, the transition occurs at a lower value 
$\Gamma_c$, the point at which the free-energies of the
 {\sc sg} and {\sc pm}$_1$ states
cross. These are respectively given by 
\begin{equation}
f_{{\sc sg}} = - \frac{1}{2\beta m} \ln(1+x_p) +\frac{\sqrt{\Gamma z}}{2} +
\frac{z}{2} (q_d-1) - \frac{p \beta m}{4x_p} q_d^p - \frac{\beta m}{4}
q_d^p
\; 
\end{equation}
and
\begin{equation}
f_{{\sc pm}_1} ={\Gamma\over 8} \; .
\end{equation}
 At $\Gamma_c$, $q_{\sc ea}$ is finite and 
$m=0$. The transition is thus of first order. 

\subsubsection{The condition of marginal stability in the static approximation}

The condition of marginal stability leads to a different equation 
for the break point $m$. In this prescription one 
does not require $m$ to be an extremum of the free-energy; instead, 
one requires that the replicon eigenvalue vanishes. 
In the static approximation, the expression derived in  
 Section~{\ref{exactmarginal} yields:
\begin{equation}
\frac{1}{(q_d-q_{\sc ea})^2}-{\beta^2 \over 2}p(p-1)q^{p-2}_{\sc ea}=0.
\label{replicao}
\end{equation}
The other extremal conditions remain unchanged. Combining 
Eqs.~(\ref{eq:q1eq_static_sg}) and (\ref{replicao}), and using the 
definitions (\ref{eq:2par}) we arrive at    
\begin{equation}
m=(p-2) \frac{1-y}{y} \;\;\;\;\;\; \Leftrightarrow \;\;\;\;\;\; x_p=p-2.
\end{equation}

For reasons that will become apparent in Section \ref{real_time}, we  define the {\it dynamic}
 transition line as the boundary of the region in
the $T-\Gamma$ plane, with $T<T^{\star}$,  where the marginally stable 
{\sc sg} exists.
The line $\Gamma_d(T)$ consists of two sections. 
The first one, that starts at the classical critical point, $(T_d,0)$,
is determined by the condition $m=1$. The dynamic transition 
in this case is of the same nature as in the classical case.  The condition $m=1$ 
can be fulfilled  until reaching a tricritical point $(T_d^{\star},\Gamma_d^{\star})$.
At lower temperatures the marginal solution disappears before
 reaching the value $m=1$.

 The dynamic phase diagram for $p=3$ is
shown in Fig.\,\ref{fig:phasedia} (thick lines).
As in the equilibrium case, $m$ is
discontinuous across the dashed line. $\Gamma_d$ lies always
above $\Gamma_{c}$. According to the interpretation of this line
given in Section \ref{real_time}, this means that the equilibrium state can
never be reached dynamically starting from an initial state in the {\sc pm}
phase.
The two lines come very close to each
other for $T \sim T^{\star}$. Within the accuracy of our calculations 
we cannot assert whether they precisely touch at $T^{\star}$, an intriguing 
possibility.
For $T < T^{\star}$, $m$ varies continuously along
$\Gamma_d(T)$ and vanishes at the quantum
critical point.

It is interesting to note that the $180^o$ rule \cite{Wheeler} does not hold in this out of equilibrium 
situation. Indeed, in Fig~\ref{fig:phasedia} one sees that the first order and 
second order line join continuously at the tricritical point.

\subsubsection{The marginal stable solution in the zero temperature limit}

Again in this case it is simple to obtain an analytic equation 
for $q_d(\Gamma)$ at 
zero temperature. 
Inserting $x_p=p-2$ in Eq.~(\ref{eq:Gamma_qd_eq_T0})
leads to  
\begin{equation}
\Gamma = \sqrt{\frac{8 p(p-2)^2}{p-1} } (1-q_d)^2 q_d^{(p-2)/2}  
\; .
\end{equation} 
The  maximum of this curve is also located at $q^{\sc max}_d=(p-2)/(p+2)$
and the physically meaningful solution is on the branch to the right of the maximum. 
This equation has a solution up to
\begin{equation}
\Gamma^{\sc marg}_{\sc max}=
\sqrt{\frac{8 p(p-2)^2}{p-1} } 
\left(\frac{p-2}{p+2}\right)^{(p-2)/2}
\left(\frac{4}{p+2}\right)^2
\; .
\label{eq:margmax}
\end{equation}
At zero temperature,  the 1-step {\sc rsb} solution with the marginal 
condition disappears at
$\Gamma=\Gamma^{\sc marg}_{\sc max}$ where $m$ is zero. 

Note that if we compare the maximum value of $\Gamma$ for which 
there is a 1-step {\sc rsb}  solution with the equilibrium and marginal criteria we have 
\begin{equation}
\Gamma^{\sc static}_{\sc max} > \Gamma^{\sc marg}_{\sc max} . 
\label{inequality}
\end{equation}
Hence the marginal solutions disappear at lower values
of $\Gamma$ than the equilibrium solutions. This result is also obtained with the quantum TAP approach \cite{Bicu} when one follows 
metastable states as functions of their energy density. 
This behavior is similar
to that of the classical problem with varying temperature: the equilibrium 
solution disappears at higher temperatures than the marginal 
solutions \cite{Kupavi}. This does not mean that
 the static transition occurs at a 
higher value of $\Gamma$  at fixed temperature. In fact, the static transition 
is determined by the identity of the free-energies $f_{\sc sg}=f_{{\sc pm}_1}$ 
that at zero temperature yields  $\Gamma_{c}^{\sc static} \simeq 3.61$ and it 
is smaller than the dynamical critical value 
$\Gamma_c^{\sc dyn} \equiv \Gamma_{\sc max}^{\sc marg}
\simeq 3.97$.

The inequality~(\ref{inequality}) is valid at finite temperatures 
and beyond the static approximation, too. 
This can be checked via the full numerical solution of the problem or with 
the quantum {\sc tap} approach used in Ref. \cite{Bicu}.

\subsection{The simpler {\bf $p=2$} case}
\label{p2_sect}

The model with $p=2$ has been studied by several authors
\cite{p2,Sachdev} and already at the classical limit it is known
to show important differences with respect to the case $p\geq 3$. The
transition is always second order and it is accompanied by a divergence
of the {\sc sg} susceptibility. 
There is no multiplicity of fully stable paramagnetic solutions and 
the spin-glass phase is RS. This statement follows from the fact that,
 for $p=2$, 
  Eq.~(\ref{eq:xxx}) only admits $x_2=0$ as a solution. Then,
Eq.~(\ref{eq:2par}) implies that, for $q_{\sc ea}\ne 0$, $m\equiv 0$.
Notice that, if $p=2$, the replicon eigenvalue (\ref{eq:repstat}) is 
identically zero.
The {\sc rs} solution is thus marginally stable, as in the 
classical case \cite{Jokoth}.

\section{The exact solution}
\setcounter{equation}{0}
\renewcommand{\theequation}{\thesection.\arabic{equation}}
\label{exact}

The static approximation neglects the 
imaginary-time dependence of the diagonal elements of the
order-parameter matrix. 
In this Section we discuss the exact equations for the paramagnetic
and spin-glass phases. We study the properties of the {\sc sg} phase using
both 
equilibrium and marginality conditions. In the latter case, we derive exact 
equations for the Edwards-Anderson parameter and the 
break point $m$ that we shall later compare to the results of the
 real-time calculation
(Section~\ref{real_time}). We present a low-temperature and low-frequency
approximation that allows us to show that the marginally stable state 
 is the only gapless solution. It also allows us to compute the 
 dependence of $\tilde q_d(0)$ on $\Gamma$, 
the $T$-dependence of the specific heat, and the low-temperature properties of $\chi(\omega\to 0)$.
 
\subsection{Equations for the exact paramagnetic solution}
\label{exactpm}
In the paramagnetic phase the order parameter matrix is diagonal,
\begin{equation}
{\tilde{Q}}_{ab}(\omega_k) = {\tilde{q}}_d(\omega_k) \delta_{ab}
\; .
\end{equation}
Replacing in Eq.~(\ref{eq:ft}) we obtain  the following equation
for $\tilde q_d(\omega_k)$
\begin{equation}
\frac{\omega_k^2}{\Gamma} + z  =
{1\over {\tilde{q}}_d (\omega_k)}+\tilde{\Sigma}(\omega_k)
\; ,
\label{eq:rspm}
\end{equation}
where we have defined the self-energy
\begin{equation}
\tilde \Sigma(\omega_k) \equiv {p\over 2}
\int_{0}^{\beta}d\tau \, \exp(i\omega_k\tau) \, 
q_d^{p-1}(\tau)
\; .
\label{eq:sigma_def}
\end{equation}
The Lagrange multiplier $z$ is determined by the spherical constraint 
\begin{equation}
q_d(0)={1\over \beta}\sum_{k}{\tilde{q}}_d(\omega_k) = 1 \; .
\label{eq:sc1}
\end{equation}
Thus, in the general case, $z$ is determined by a non-trivial implicit
equation in contrast with its
analog in the static approximation, 
Eq.~(\ref{eq:z_static_pm}). It follows from Eq.~(\ref{eq:rspm}) 
that
the Fourier components of $q_d(\tau)$ with $k \ne 0$ are strongly
suppressed  in the limit $\Gamma \to 0$. The normalization condition
then implies that 
$\lim_{\Gamma \to 0} \tilde q_d(\omega_k) = \beta \delta_{\omega_k,0}$ 
which is the classical result.

\subsection{Equations for the exact spin-glass solution}
\label{exactsg}
The {\sc rs} {\it Ansatz} in Fourier space is given by
\begin{equation}
{\tilde{Q}}_{ab}(\omega_k) = \left({\tilde{q}}_d(\omega_k)
-{\tilde{q}}_{\sc ea}\right)\delta_{ab} + {\tilde{q}}_{\sc ea}
\; .
\label{eq:rs_sol}
\end{equation}
The inverse matrix, ${\tilde{\bf Q}}^{-1}$, 
can be easily evaluated to yield (in the limit $n\to 0$)
\begin{equation}
\left({\tilde{\bf Q}}^{-1}\right)_{ab}(\omega_k) =
{1\over {\tilde{q}}_d(\omega_k)-{\tilde{q}}_{\sc ea}}\delta_{ab}
-{{\tilde{q}}_{\sc ea}\over 
\left({\tilde{q}}_d(\omega_k)-{\tilde{q}}_{\sc ea}\right)^2}
\; .
\end{equation}
Inserting the above expressions in Eq.~(\ref{eq:ft}) 
and using the identity (\ref{eq:offd}), we get 
(for $a\ne b$ and $\omega_k = 0$)
\begin{equation}
-{q_{\sc ea}\over \left({\tilde{q}}_d(0)- \beta q_{\sc ea} \right)^2}
+{p\over 2}q^{p-1}_{\sc ea}=0
\; .
\label{eq:rs}
\end{equation}
Note that in the classical limit $\tilde{q_d}(0)=\beta$, 
and one recovers the classical equation 
for $q_{\sc ea}$ in the {\sc rs} case \cite{crisanti}. 
As in the static approximation, 
the replica symmetric {\it Ansatz} is unstable towards replica
symmetry breaking. The replicon 
eigenvalue associated with Eq.~(\ref{eq:rs_sol}) is
\begin{equation}
\Lambda_T=\frac{\beta^2}{({\tilde{q}}_d(0)-\beta q_{\sc ea})^2}
-\frac{\beta^2}{2} p(p-1)q_{\sc ea}^{p-2}
\;.
\end{equation}
This equation and Eq.~(\ref{eq:rs}) yield $\Lambda_T=p(2 - p)q_{\sc
EA}^{(p-2)}/2$. This eigenvalue is negative for all $p\ge 3$. This  
excludes a {\sc rs} stable {\sc sg} state and the symmetry between the replicas must be
broken. 

Inspired by the classical case \cite{crisanti}, 
and the results from the static approximation, 
we use a 1-step {\sc rsb} {\it Ansatz} that we prove to be exact in the 
Appendix.
In Fourier space this can be written in the form
\begin{equation}
{\tilde{Q}}_{ab}(\omega_k)=\left({\tilde{q}}_d(\omega_k)
-{\tilde{q}}_{\sc ea}\right)\delta_{ab}+
\left({\tilde{q}}_{\sc ea} - 
{\tilde{q}}_0\right)\epsilon_{ab} + {\tilde{q}}_0
\; ,
\label{eq:rsb1}
\end{equation}
where the matrix ${\bf \epsilon}$ has been  defined in
Eq.~(\ref{eq:eps_ab}).
In the absence of an external field, 
the saddle point equation for $q_0$ 
yields $q_0 = 0$, as in the static approximation and the classical case \cite{crisanti}.
The inverse matrix, ${\tilde {\bf Q} }^{-1}(\omega_k)$ can then 
be written as
\begin{equation}
\left({\tilde{Q}}^{-1}\right)_{ab}(\omega_k) = 
A(\omega_k)\delta_{ab}+B(\omega_k)\epsilon_{ab}
\; ,
\label{eq:rsb2}
\end{equation}
where (in the limit $n\to 0)$
\begin{equation}
A(\omega_k)= 
{1\over 
{\tilde{q}}_d(\omega_k)- {\tilde{q}}_{\sc ea}}
\label{eq:a}
\end{equation}
and
\begin{equation}
B(\omega_k)=
-{{\tilde{q}}_{\sc ea}\over
{\tilde{q}}_d^2(\omega_k)- {\tilde{q}}_{\sc ea}^2(m-1)+
{\tilde{q}}_d(\omega_k){\tilde{q}}_{\sc ea}(m-2)}
\; .
\label{eq:b}
\end{equation}
The saddle point equations for $q_{\sc ea}$ and ${\tilde{q}}_d(\omega_k)
$ can be obtained by inserting Eqs.~(\ref{eq:rsb1}) and 
(\ref{eq:rsb2}) in Eq.~(\ref{eq:ft}) at $a\ne b, \omega_k=0$ 
and $a=b$, respectively. They read
\begin{equation}
-{1\over
{\tilde{q}}_d^2(0)-\beta^2q_{\sc ea}^2(m-1)+\beta
q_{\sc ea}{\tilde{q}}_d(0)(m-2)}+
{p\over 2}q_{\sc ea}^{p-2}=0\; ,
\label{eq:q1}
\end{equation}
and
\begin{equation}
\frac{\omega_k^2}{\Gamma}+ z 
= 
{{\tilde{q}}_d(\omega_k)+\beta q_{\sc ea}(m-2)\delta_{\omega_k,0}
\over
{\tilde{q}}_d^2(\omega_k)-\beta^2q_{\sc ea}^2(m-1)\delta_{\omega_k,0}
+\beta q_{\sc ea}{\tilde{q}}_d(\omega_k)(m-2)\delta_{\omega_k,0}}
+ \tilde \Sigma(\omega_k)
\; ,
\label{eq:qw}
\end{equation}
with $\tilde\Sigma(\omega_k)$ defined in Eq.~(\ref{eq:sigma_def}).
Equations~(\ref{eq:q1}) and (\ref{eq:qw}) must be supplemented by an
equation for the break point parameter, $m$. As within the static
approximation, the latter may be obtained using two different
prescriptions 
that we discuss next.

\subsection{The equilibrium spin-glass solution}
\label{exactequi}

In equilibrium, the free-energy per spin, $f$, must be stationary
with respect to variations in $m$. 
The only $m$-dependent terms in $f$ are 
\begin{equation}
{1\over \beta}\lim_{n\to 0}{1\over n}
\left(-{1\over 2}\mbox{Tr}\ln\left(\beta^{-1} {\tilde{\bf Q}}(0)\right)-
{\beta^2\over 4}n(m-1)q_{\sc ea}^p\right)
\label{eq:fe2}
\end{equation}
which can be easily evaluated with the result
\begin{equation}
-{1\over 2\beta}\left[\ln\left(\frac{{\tilde{q}}_d(0)-
{\tilde{q}}_{\sc ea}(1-m)}{\beta}\right)+{m-1\over m}\ln\left(
{{\tilde{q}}_d(0)-{\tilde{q}}_{\sc ea}\over {\tilde{q}}_d(0)-
{\tilde{q}}_{\sc ea}(1-m)}\right)\right]
-{\beta\over 4}
(m-1)q_{\sc ea}^p.
\label{eq:fe3}
\end{equation}
The extremization equation $\partial f
/\partial m =0$ is 
\begin{equation}
{1\over m}{\beta q_{\sc ea}\over 
{\tilde{q}}_d(0)-\beta q_{\sc ea} (1-m)}+{1\over m^2}\ln\left(
{{\tilde{q}}_d(0)-\beta q_{\sc ea}\over 
{\tilde{q}}_d(0)-\beta q_{\sc ea} (1-m)}\right)+
{\beta^2\over 2}q_{\sc ea}^p=0.
\label{eq:m}
\end{equation}
Combining Eqs.~(\ref{eq:q1}) and (\ref{eq:m}) and defining
\begin{equation}
y'=\frac{\beta q_{\sc ea}}{{\tilde{q}}_d(0)}
\;\;\;\;\;\;\;\;\,\,\,\,\,\,\,
x_p={ m y'\over 1-y'},
\label{defxy}
\end{equation}
we obtain the same expression for $x_p$ that we derived in the static
approximation, Eq.~(\ref{eq:xxx}).  Using Eq.~(\ref{eq:q1}) and the 
definitions 
(\ref{defxy}), we have 
\begin{equation}
{p(\beta m)^2q_{\sc ea}^p\over 2}={2x_p^2\over 1+x_p}
\label{eq:q1p}
\end{equation}
and
\begin{equation}
{p(\beta m)^2{\tilde{q}}_d^p(0)\over 2}=\beta^p x_p^{2-p}
{(m+x_p)^p\over 1+x_p}
\; .
\label{eq:qdp}
\end{equation}
The first of these equations is identical 
to Eq.~(\ref{eq:q1_x_stat}).

It is convenient to separate $q_d$ and the self-energy into constant
and $\tau$-dependent parts,
\begin{eqnarray}
\label{def_reg}
q_d(\tau) &=& q_{\sc ea} + q_{\sc reg}(\tau)
\; ,
\\
\Sigma(\tau)&=&\frac{p}{2} q_{\sc ea}^{p-1} + \Sigma_{\sc reg}(\tau)
\; .
\end{eqnarray}
Substituting in Eq.~(\ref{eq:qw}) and using Eqs.~(\ref{eq:q1p}) 
and (\ref{eq:qdp}),  the terms
that are proportional to $\delta_{\omega_k,0}$ cancel and we obtain an
equation for the regular part of $\tilde{q_d}(\omega_k)$, 
\begin{equation}
\label{chireg}
\frac{\omega_k^2}{\Gamma} + z' = 
\frac{1}{\tilde q_{\sc reg}(\omega_k)} +
\tilde{\Sigma}_{\sc reg}(\omega_k)- \tilde{\Sigma}_{\sc reg}(0)
\; ,
\end{equation}
where 
\begin{eqnarray}
\label{z'}
z'&=&\frac{p}{2}\  \beta m q_{\sc ea}^{p-1} \frac{1+x_p}{x_p}
\end{eqnarray}
and
\begin{equation}
\tilde\Sigma_{\sc reg}(\omega_k) - \tilde\Sigma_{\sc reg}(0) =
\frac{p}{2} \int_0^\beta d\tau \; (\cos(\omega_k\tau)-1) \, 
(q_d^{p-1}(\tau)-q_{\sc ea}^{p-1})
\; .
\label{eq:diff1}
\end{equation}

\subsection{The marginally stable spin-glass solution}
\label{exactmarginal}

In this Section we compute the replicon eigenvalue for the 
exact problem and we derive the consequences of using the 
condition of marginal stability that corresponds to setting it
to zero. 

\subsubsection{The replicon eigenvalue}
\label{derivation_replicon}
To derive the replicon eigenvalue, we first 
compute the second order variation of $G_0$ with 
respect to ${\tilde{X}}_{ab}(\omega_k)\equiv {\tilde{Q}}_{ab}(\omega_k)/\beta$ which gives 
the Gaussian ${\tilde{\bf X}}$ fluctuations:
\begin{eqnarray}
2\delta^2 G_0 & =& \sum_{\omega_k} \mbox{Tr}\left(
{\tilde{\bf X}}^{-1}(\omega_k)\cdot   \delta {\tilde{\bf X}}
(\omega_k)\right)^2
\nonumber
\\
& & 
-{\beta \over 2}p(p-1)\sum_{ab}\sum_{
\omega_k'}\sum_{\omega_k''}\int d\tau \left(
\sum_{\omega_k}\exp (-i\omega_k\tau){\tilde{X}}_{ab}(\omega_k)
\right)^{p-2}
\nonumber\\& &\times   \exp\left[-i\tau\left(\omega_k'+\omega_k''\right)
\right]\delta{\tilde{X}}_{ab}(\omega_k')
\delta{\tilde{X}}_{ab}(\omega_k'')
\; .
\label{eq:g02}
\end{eqnarray}
For the 1-step {\sc rsb} {\it Ansatz} the eigenvalues and eigenvectors
of this 
quadratic form may be obtained by solving the equation 
\begin{eqnarray}
A^2(\omega_k)\delta{\tilde{X}}_{ab}(\omega_k)+
A(\omega_k)B(\omega_k)\left[\left(\delta{\tilde{\bf X}}
(\omega_k)\cdot   {\bf \epsilon}\right)_{ab}+\left(
{\bf \epsilon}\cdot   \delta{\tilde{\bf X}}(\omega_k)\right)_{ab}
\right]
\nonumber\\+ B^2(\omega_k)\left(\epsilon\cdot   \delta{\tilde{\bf X}}
(\omega_k)\cdot   {\bf \epsilon}\right)_{ab}
-{p(p-1) \over 2} q_{\sc ea}^{p-2}\epsilon_{ab}
\delta {\tilde{X}}_{ab}(\omega_k)=\beta^{-2} \Lambda_T (\omega_k)
\delta {\tilde{X}}_{ab}(\omega_k)
\; ,
\label{eq:ev}
\end{eqnarray}
where $a\ne b$ and $A(\omega_k)$ and 
$B(\omega_k)$ are defined by Eqs.~(\ref{eq:a}) and 
(\ref{eq:b}). Since we are looking for the 
replicon (transverse) eigenvalue we impose the constraints
\begin{equation}
\left({\bf \epsilon}\cdot   \delta {\tilde{\bf X}}
(\omega_k)\right)_{ab} = 0 \;,
\end{equation}
\begin{equation}
\left(1-\epsilon_{ab}\right)\delta{\tilde{X}}_{ab}
(\omega_k) = 0
\; ,
\end{equation}
and set $\omega_k = 0$ to obtain
\begin{equation}
\Lambda_T = \beta^2 A^2(0)-
{\beta^2 \over 2}p(p-1)q_{\sc ea}^{p-2}.
\end{equation}
The condition of marginal stability corresponds to 
$\Lambda_T=0$ and replaces (\ref{eq:m}) for the determination of the
break point.

\subsubsection{The block-size $m$}

Setting $\Lambda_T=0$
 and using Eq.~(\ref{eq:q1}) we obtain
\begin{equation}
m=(p-2){{\tilde{q}}_d(0)-\beta q_{\sc ea} \over 
\beta q_{\sc ea}}= (p-2)\frac{1-y'}{y'}.
\label{eq:rep}
\end{equation}
It follows from this expression and Eq.~(\ref{eq:qdp}) 
that $x_p = p-2$.  This is the result that we obtained 
in the static approximation. 
In the classical limit, $m=(p-2)(1-q_{\sc ea})/q_{\sc ea}$, a known
result \cite{crisanti} . Moreover, using again Eq.~(\ref{eq:q1}), the above expression 
can be written as
\begin{equation}
\beta m=(p-2)\sqrt{{2\over p(p-1)}}
q_{\sc ea}^{-p/2}
\; , 
\label{eq:rep2}
\end{equation}
which is identical to Eq.~(\ref{eq:q1_x_stat}) obtained in the static approximation. 

Equations (\ref{chireg})-(\ref{eq:diff1}), derived for the equilibrium
state, also hold for the marginal {\sc sg} provided we make the
substitution $x_p \to p-2$. 

\subsubsection{The Edwards-Anderson order parameter}

If we rewrite the denominator in the first term of
Eq.~(\ref{eq:q1}) and replace $m$ by its expression in 
Eq.~(\ref{eq:rep2}) we obtain a new equation for $q_{\sc ea}$ 
that reads
\begin{equation}
\label{handyequation}
1 = 
\frac{p(p-1)}{2} 
\left(\tilde q_d(0) - \beta q_{\sc ea}\right)^2 q_{\sc ea}^{p-2} =
\frac{p(p-1)}{2}  q_{\sc ea}^{p-2} \tilde{\chi}_{\sc reg}^2(0)
\; .
\label{eq:rep3}
\end{equation}
Note the absence of explicit dependence upon $\Gamma$ 
both  here and in Eq.~(\ref{eq:rep2}). The quantum parameter only  
enters implicitly through $\tilde q_d(0)$ and $ q_{\sc ea}$. 

An identical equation for the Edwards-Anderson parameter 
can also be derived with the quantum TAP approach~\cite{Bicu}
when the threshold value for the free-energy density is chosen.  
In the classical limit, $\tilde q_d(0)=\beta$ and  
this equation becomes the equation for $q_{\sc ea}$
of the classical TAP states of the threshold level~\cite{Kupavi}. 
 
This equation may be studied in the usual way, {\it i.e.}, 
by looking at the shape of the function that 
appears on its right-hand-side.
This vanishes at
$q_{\sc ea}=0$ and $q_{\sc ea}=\tilde q_d(0)/\beta$ and has a maximum 
at $\beta q_{\sc ea}=\frac{p-2}{p} \tilde q_d(0)$. 
The maximum is at $m=2$ just as in 
the classical problem. Two solutions for $\beta q_{\sc ea}$, one on the 
right-branch, the other one on the left-branch.
It is simple to see that the correct physical behavior 
of a decreasing Edwards-Anderson parameter with increasing 
temperature is achieved by the solution on the right-branch.

Note, however, that other equations where 
$\tilde q_d(0)$ appears
need also be satisfied in the quantum problem. These 
depend explicitly on the quantum parameter $\Gamma$. 
Depending on $\Gamma$ and $T$ the marginally stable {\sc sg} 
may disappear before reaching the $m=2$ value. 

\subsubsection{The low-frequency limit}
\label{approx_exact}

The existence of a vanishing transverse eigenvalue implies that 
the spectrum of magnetic excitations of the marginally 
stable {\sc sg} state is gapless. Therefore, at $T=0$, 
$q_{\sc reg}(\tau)$ is expected to decay asymptotically as a  power-law:
\begin{equation}
q_{\sc reg}(\tau) \sim \frac{A}{|\tau|^\alpha}
\;\;\;\;\;
|\tau|\to\infty
\; .
\label{asym}
\end{equation}
Equivalently, the imaginary part of the susceptibility $\chi''(\omega)$ 
 vanishes as $\omega^{(\alpha
-1)}$ as $\omega \to 0$.
In the zero temperature limit the Matsubara frequencies are continuous
and Eq.~(\ref{eq:diff1}) becomes
\begin{equation}
\label{sigomega}
\tilde{\Sigma}_{\sc reg}(\omega) -
\tilde{\Sigma}_{\sc reg}(0)=\frac{p}{2}\int_{-\infty}^{\infty} d\tau \ \left(\cos\omega
\tau - 1\right) \left[(p-1) q_{\sc ea}^{p-2} q_{\sc reg}(\tau) + \ldots \right]
\; .
\label{eq:diff2}
\end{equation}
The terms represented by the dots contain higher powers of
$q_{\sc reg}(\tau)$. In the $\omega\to 0$ the 
integral is dominated by the 
long $\tau$ limit, and we may replace $q_{\sc reg}(\tau)$ by its
asymptotic form (\ref{asym}). Eq.~(\ref{eq:diff2}) becomes
\begin{equation}
\tilde{\Sigma}_{\sc reg}(\omega) -
\tilde{\Sigma}_{\sc reg}(0)\approx
\frac{p(p-1)}{2} A q_{\sc ea}^{p-2} \int_{-\infty}^{\infty} d\tau
\; 
\frac{\cos(\omega \tau)-1}{|\tau|^\alpha} 
\; ,
\end{equation}
leading to
\begin{equation}
\tilde{\Sigma}_{\sc reg}(\omega) -
\tilde{\Sigma}_{\sc reg}(0)\approx 
p(p-1) A q_{\sc ea}^{p-2} \Gamma[1-\alpha] \, \sin\left(\frac{\alpha\pi}{2}\right) \, 
|\omega|^{\alpha-1}
\; ,
\label{sig_approx}
\end{equation}
where $\Gamma[x]$ is the Gamma function. The higher order terms
neglected in Eq.~(\ref{eq:diff2}) give rise to terms vanishing as
higher powers of $\omega$.  Likewise, 
\begin{equation}
\tilde q_{\sc reg}(\omega) -
\tilde q_{\sc reg}(0)\approx
2 A  \Gamma[1-\alpha] \, \sin\left(\frac{\alpha\pi}{2}\right) \, 
|\omega|^{\alpha-1} \; .
\label{q_approx}
\end{equation}
Replacing Eqs.~(\ref{sig_approx}) and (\ref{q_approx}) in
Eq.~(\ref{chireg}) and  comparing terms of order $|\omega|^{\alpha-1}$
we obtain
\begin{equation}
z' = \frac{p(p-1)}{2} \, q_{\sc ea}^{p-1} \beta \frac{1-y'}{y'}
\; ,
\end{equation}
which compared with Eq.~(\ref{z'}) gives 
\begin{equation}
x_p=p-2
\; ,
\label{xpmar}
\end{equation}
the marginal value of $x_p$. We thus see that the marginally stable
state is the only one compatible with a gapless spectrum. Conversely,
the equilibrium state must necessarily have a gap in its excitation spectrum.

In order to determine the values of  the exponent
$\alpha$ and the amplitude $A$, we proceed by keeping only the first 
term on the rhs of Eq.~(\ref{eq:diff2}), 
\begin{equation}
\label{approx}
\tilde{\Sigma}_{\sc reg}(\omega_k) -
\tilde{\Sigma}_{\sc reg}(0)\approx \frac{p (p-1)}{2} q_{\sc ea}^{p-2}
\left(\tilde q_{\sc reg}(\omega_k) - \tilde q_{\sc reg}(0)\right)
\; ,
\label{eq:diff3}
\end{equation}
and replace it in Eq.~(\ref{chireg}) noting that  $\tilde q_{\sc reg}(0) =
 \tilde q_d(0) - \beta q_{\sc ea} \equiv \beta m q_{\sc ea}/(p-2)$
 (cf. Eqs.~(\ref{defxy}) and (\ref{xpmar})). The result is
\begin{equation}
\label{quadratic}
\left[\frac{\omega_k^2}{\Gamma} + \frac{p (p -1)}{p-2} \beta m
q_{\sc ea}^{p-1}-\frac{p (p-1)}{2} q_{\sc ea}^{p-2} \tilde q_{\sc reg}(\omega_k)\right]
\tilde q_{\sc reg}(\omega_k)=1
\; .
\end{equation}
The solution of this quadratic equation is 
\begin{equation}
\label{quad}
\tilde q_{\sc reg}(\omega_k)=\frac{\frac{\omega_k^2}{\Gamma} + 
\frac{p (p - 1)}{p-2}
\beta m q_{\sc ea}^{p-1}
-\left|\omega_k\right| \Gamma^{-1/2}
\sqrt{\frac{\omega_k^2}{\Gamma} + \frac{2 p(p -1)}{p-2}
\beta m q_{\sc ea}^{p-1}}}{p (p-1) q_{\sc ea}^{p-2}}
\; . 
\end{equation}
This result is only valid in the limit $\omega_k \to 0$ where it reduces to 
\begin{equation}
\tilde q_{\sc reg}(\omega_k) \sim \tilde q_d(0)-\beta q_{\sc ea} - 
\frac{(\tilde q_d(0)-\beta q_{\sc ea})^{1/2} q_{\sc ea}^{1-p/2} 
\sqrt{2/[p(p-1)]}}{\sqrt{\Gamma}} \; |\omega_k|
\label{low_w}
\end{equation}
Thus, the exponent giving the
asymptotic decay of $q_d(\tau)$ in Eq.~(\ref{asym}) is $\alpha=2$. 
Substituting Eqs.~(\ref{eq:q1p}), (\ref{eq:qdp}) and (\ref{xpmar}) 
in Eq.~(\ref{low_w}) and analytically continuing to real frequencies
we obtain the exact result   
\begin{equation}
\label{eq:spectrum}
\lim_{\omega\to 0} {\chi''(\omega) \over \omega} =
{1 \over \sqrt{\Gamma}} \left[\frac{2\,q_{\rm EA}^{(2-p)}}{p (p-1)}\right]^{3/4}
\; .
\end{equation}
One can easily show that the amplitude $A$ is obtained by 
multiplying 
the rhs of the equation above  by $\pi$. 

It is interesting to notice that a linear excitation spectrum has also
been found in the case of the SU(${\cal N}$) Heisenberg {\sc sg} model
\cite{antoine}.
However, in our model this gapless spectrum is not a 
consequence of Goldstone's theorem as the model does not 
 possess any continuous symmetry.

\subsubsection{An approximate solution of the dynamical equations}
\label{approx_sol}

The results of the previous section suggest a simple 
approximate solution of the complete set of equations that becomes
exact in the low-frequency and low-temperature limit.   
Once $\tilde{q_d}(\omega)$ is known, the equation of state,
 $q_{\sc ea}(T,\Gamma)$, must be determined from the normalization
condition
\begin{equation}
\label{norm}
1-q_{\sc ea} = \frac{1}{\beta} \sum_{\omega_k} 
\tilde q_{\sc reg}(\omega_k)\equiv
\int_0^\infty \frac{d\omega}{\pi} \chi''(\omega) 
\coth\left(\beta\;\omega/2\right)
\; .
\end{equation}
If we assume that the integral is dominated by the low frequencies, we
can compute it by replacing  $\chi''(\omega)$ by the analytic continuation of
Eq.~(\ref{quad}).
The result at $T=0$ is
\begin{equation}
\label{norm1}
1-q_{\sc ea} = \frac{4}{3 \pi} \frac{\sqrt{\Gamma}}{\left[ p (p - 1)
q_{\sc ea}^{p-2}/2\right]^{1/4}} \;, 
\end{equation}
which reduces to
\begin{equation}
\Gamma = \left(\frac{3 \pi}{4}\right)^2 \left[p (p - 1)
q_{\sc ea}^{p-2}/2\right]^{1/2} (1 - q_{\sc ea})^2.
\end{equation}
This equation has the same form as the corresponding one obtained with the
static approximation, Eq.~(\ref{eq:margmax}); only the coefficient is different. 
For $p=3$, we find that the maximum coupling for the marginal {\sc
sg} state, that is also the critical coupling at zero temperature,  
is $\Gamma_d(0)=2.75$ 
compared to $\Gamma_d(0)$=3.97 in the static approximation and the 
estimate $\Gamma_d(0)=3.1$ from the numerical calculations to be
described below. 
It is easy to check that, at finite but low temperature, the
corrections to Eq.~(\ref{norm1}) are of ${\cal O}(T^2)$. Therefore,
\begin{equation}
\label{eq:q1T}
q_{\sc ea}(T, \Gamma)=q_{\sc ea}(0,\Gamma)\left[1 - {\cal O}(T^2)\right] \; ,
\end{equation}
as $T \to 0$.

 We shall compare quantitatively the low-frequency
approximation with the numerical solution of the exact equations in
Section~\ref{numerical}.

\subsection{Thermodynamic functions}
\label{thermo_functions}

The free-energy per spin may be calculated by substituting our  {\it
Ansatz} for the order-parameter matrix in Eqs.~(\ref{eq:go}) 
and  (\ref{eq:fe1}). After some algebra we obtain:
\begin{eqnarray}
\label{betaf}
\nonumber
\beta f &=&-\frac{1}{2} \left\{\log\left[1 - (1-m) y'\right] +\frac{m-1}{m}
\log\frac{1-y'}{1-(1-m) y'}\right\} \\
&-& \frac{\beta^2}{4} (m-1) q_{\rm EA}^p - \frac{\beta}{4}\int_0^{\beta} d\tau
q_d^p(\tau) - \frac{\beta z}{2} +  \ln \left[ 2 \sinh\left(\beta
\sqrt{\Gamma z}/2\right)\right]\\
\nonumber
&-&\frac{1}{2} \sum_{\omega}
\ln\left[\left(\frac{\omega^2}{\Gamma}+z\right)\tilde{q}_d(\omega)\right]
+ \frac{1}{2}  \sum_{\omega} 
\left[\left(\frac{\omega^2}{\Gamma}+z\right)\tilde{q}_d(\omega)-1\right].
\end{eqnarray}
The entropy, internal energy and specific heat are given by
\begin{eqnarray}
\label{UCS}
S=-\frac{\partial f}{\partial T} \; ,\;\;\;\;\;\;  
U=\frac{\partial (\beta
f)}{\partial \beta} \; , \;\;\;\;\;\;  
C_v = \frac{\partial U}{\partial T}
\;,
\end{eqnarray}
respectively.

The derivatives needed to compute $S$ and $U$ must be taken only with
respect to the {\it explicit} $T$-dependence of the free-energy. Contributions 
from terms of the form $\partial (\beta f)/\partial q_{\sc ea} \times 
\partial q_{\sc ea}/\partial T$ and alike vanish because of the
stationarity condition. A convenient way to proceed consists in
observing that $q_d(\tau)$ is a dimensionless
function and, as such, it
can only depend on dimensionless variables. We can choose them as follows
\begin{equation}
q_d(\tau,\tilde J,\Gamma,T)=\hat{q}_d(\tau/\beta,\beta z,\beta \tilde J,\Gamma/\tilde J) 
\; .
\label{scaling}
\end{equation}
It follows that 
\begin{equation}
U_2 \equiv-\frac{\beta}{4} \int_0^\beta d\tau q_d^p(\tau)=
 -\frac{\beta^2}{4} \int_0^1 ds\ 
\hat{q}_d^p(s,\beta z,\beta \tilde J,\Gamma/\tilde J)
\; ,
\end{equation}
and its explicit derivative with respect to
$\beta$ is 
\begin{equation}
\label{right}
\frac{\partial U_2}{\partial \beta}
=-\frac{\beta}{2}\int_0^{1} ds \hat{q}_d^p(s,\beta z,\beta 
\tilde J,\Gamma/\tilde J)=-\frac{1}{2}\int_0^{\beta} d\tau
q_d^p(\tau)
\; .
\end{equation}
Similarly, 
\begin{equation}
\label{scaling2}
\tilde{q}_d(\omega_k)=\int_0^{\beta}d\tau q_d(\tau) e^{i\omega_k
\tau}=\beta \int_0^1 ds \hat{q}_d(s) e^{i 2\pi k s}\equiv \beta \hat{q}_d(k).
\end{equation}
The explicit $\beta$ dependence of 
$\tilde{q}_d(\omega)$ is thus a multiplicative
factor. Finally, we observe that $\partial \omega_k/\partial \beta =
-\omega_k/\beta$. Using these considerations, we derived the following expression
for the internal energy:
\begin{equation}
\label{finalU}
U=-\frac{\beta}{2}(m-1) q_{\rm EA}^p -\frac{1}{2}\int_0^{\beta}d\tau
q_d^p(\tau) + \frac{z}{2} + \frac{1}{2 \beta}\sum_{\omega_k} \left[1 -
\left(\frac{\omega_k^2}{\Gamma}+z\right)\tilde{q}_d(\omega_k)\right] .
\end{equation}
We analyze separately the {\sc pm} and {\sc sg} 
phases.

\subsubsection{Internal energy of the paramagnetic phase}

The equation of motion can be recast as
\begin{equation}
\label{para}
1-
\left(\frac{\omega_k^2}{\Gamma}+z\right)\tilde{q}_d(\omega_k)=-\tilde \Sigma(\omega_k)
\tilde{q}_d(\omega_k)
\; ;
\end{equation}
replacing this expression in the last term of Eq.~(\ref{finalU}) and using
\begin{equation}
\label{astuce}
\frac{1}{\beta}\sum_{\omega_k} \Sigma(\omega_k) \tilde{q}_d(\omega_n)
\equiv \int_0^{\beta} d\tau \; \Sigma(\tau) q_d(\tau) =\frac{p}{2}
\int_0^{\beta}d\tau \; q_d^p(\tau)
\; ,
\end{equation}
we obtain a very simple expression for the paramagnetic internal energy:
\begin{equation}
\label{upara}
U_{\sc pm}=\frac{z}{2}-\frac{p+2}{4} \int_0^{\beta} d\tau \; q_d^p(\tau)
\; .
\end{equation}

Since we have shown that a gapless spectrum is only possible in the
marginally stable {\sc sg} state, there must be a gap $E_{\sc
g}$ in the spectrum of the paramagnetic phase. Therefore the specific heat in the
{\sc pm} phase vanishes exponentially at low temperature:
\begin{equation}
C_v \propto \exp(-E_{\sc g}/T)
\; .
\end{equation}

\subsubsection{Internal energy in the spin-glass phase}

The same procedure can be applied to the {\sc sg} phase,
 using this time the equation of
motion in the form 
\begin{equation}
\label{SG}
\left[\frac{\omega_k^2}{\Gamma} + z - \tilde \Sigma(\omega_k)\right]
\tilde{q}_d(\omega_k)=1 + (f_p - 1) \delta_{\omega_k,0}
\; ,
\end{equation}
where 
\begin{equation}
\label{f}
f_p = 1 +\frac{x_p^2}{m (1 + x_p)} - \frac{x_p^2}{m^2 (1 + x_p)} \; .
\end{equation}
By substituting Eq.\,(\ref{SG}) into Eq.\,(\ref{finalU}) we obtain
\begin{equation}
U_{\sc sg} =\frac{z}{2} -\frac{p+2}{4} \,\beta m \,q_{\rm EA}^p - \frac{p+2}{4}
\int_0^{\beta} d\tau \left(q_d^p(\tau) - q_{\rm EA}^p\right)
\; ,
\label{internal_sg}
\end{equation}
where $\beta m$ is given by Eq.~(\ref{eq:q1p})
and $z$ 
\begin{equation}
\label{zsg}
z =\frac{p}{2} \beta m \,q_{\rm EA}^{p-1}\,\frac{1 + x_p}{x_p} + 
\tilde \Sigma_{\sc reg}(\omega=0) \; ,
\end{equation}
and 
\begin{equation}
\label{regse}
\tilde\Sigma_{\sc reg}(\omega=0)=\frac{p}{2} \int_0^{\beta} d\tau
\left(q_d^{p-1}(\tau) - q_{\rm EA}^{p-1}\right)
\; .
\end{equation}

In the equilibrium {\sc sg} state the specific heat vanishes
exponentially as its excitation spectrum has a gap.

Since $f$ is not an
extremum with respect to $m$  in the marginal state,
 it would seem that the term 
$\partial (\beta f)/\partial m \times
\partial m/\partial T$ should appear in the calculation of its internal
energy. However, it can be shown that in the correct expression of the
energy density of this state, that coincides with the one obtained 
in the real-time dynamics calculation, such
contributions are absent. 
Hence, we shall use Eq.~(\ref{finalU}) also
in the marginal case. Since now $q_{\sc reg}(\tau)\equiv
q_d(\tau)- q_{\rm EA} \sim \tau^{-1}$ at $T=0$, 
we expect $C_v$ to have a power-law dependence upon $T$. 
Indeed, since at low
temperature the integral in Eq.~(\ref{internal_sg}) is dominated by
the long-time behavior of $q_d(\tau)$, we can evaluate the latter by 
Fourier transforming Eq.~(\ref{quad}). The resulting expression
depends on $T$ only through $q_{\sc ea}$. It
follows from  Eq.~(\ref{eq:q1T}) that, in the marginal {\sc sg} state,
 $C_v \propto T$. This power-law behavior
is reminiscent of the 
temperature dependence of low-temperature glasses that has been 
intensively studied experimentally~\cite{Osheroff}.

\section{Numerical solution}
\label{numerical}
\setcounter{equation}{0}
\renewcommand{\theequation}{\thesection.\arabic{equation}}

In this Section we present numerical solutions obtained
using an algorithm that solves the equations of Section~\ref{exact}. All
the results reported in this Section where obtained for $p=3$.

\subsection{The paramagnetic phase}
\label{num_pm}

The {\sc pm} phase is described by  
Eq.~(\ref{eq:rspm}) subject to the spherical constraint (\ref{eq:sc}).  
At each temperature, the equations are solved iteratively
as a function of $\Gamma$, starting from high and low values of
this parameter. In these limits we can find analytical perturbative
solutions that are used as  starting points for the iteration procedure. 
For each value of $\Gamma$, the input is the converged
solution obtained for the previous value. 
 
\begin{figure}[ht]
\epsfxsize=3in 
\centerline{\hbox
{
\epsffile{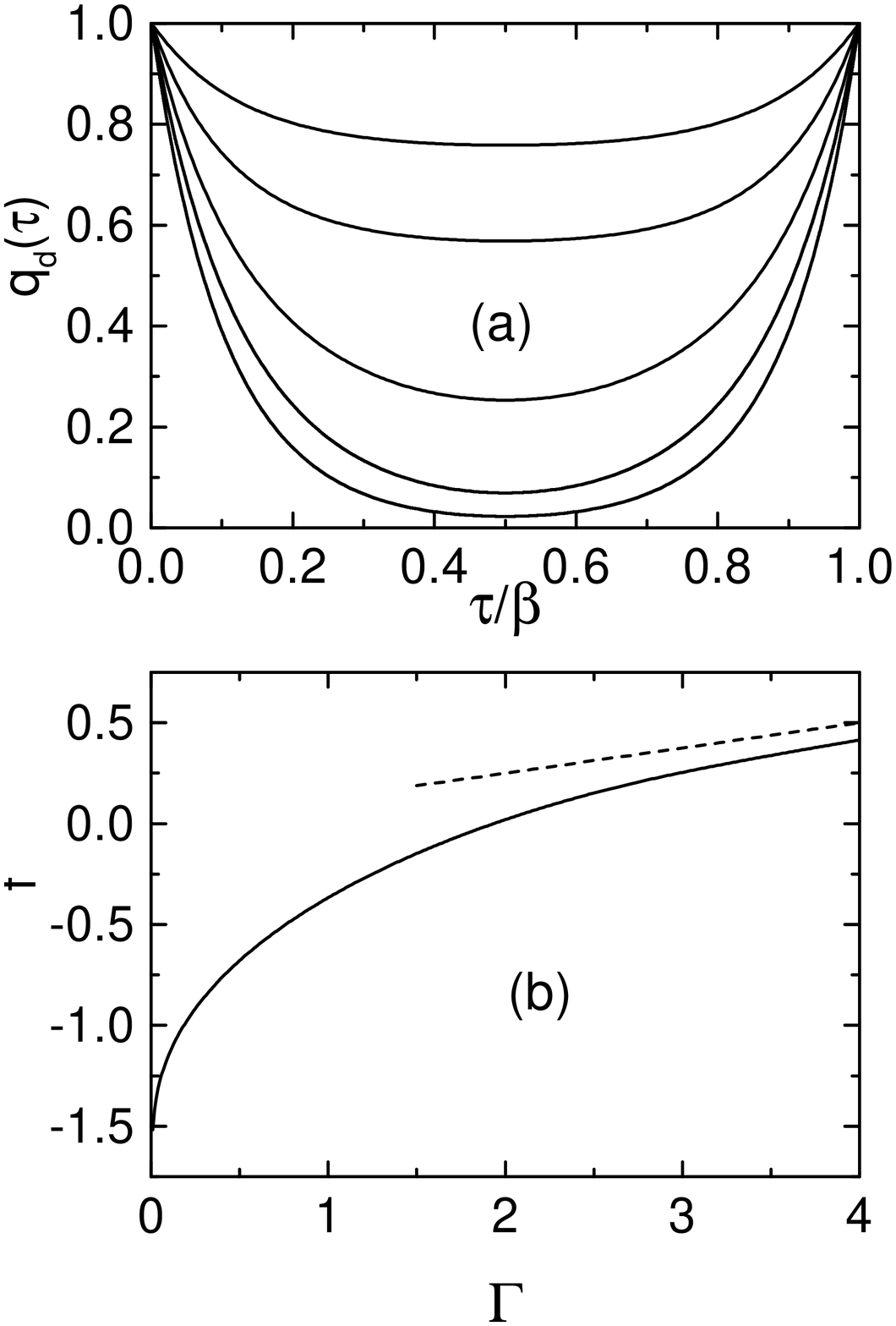}
}}
\caption{(a) The function $q_d(\tau)$ for the {\sc pm} solution at
$\beta=4$ and $\Gamma=1, 2, 3,4, 5$ from top to bottom. 
(b) The {\sc pm} free-energy as a function
of $\Gamma$ at $\beta=4$. The dashed line is the asymptotic
value $f=\Gamma/8$.}
\label{qtb4}
\end{figure}

For $\beta < \beta_{\sc p}\approx 6$,
Eq.~(\ref{eq:rspm}) has only one solution that is connected
continuously to both the high and  low-$\Gamma$ limits. 
Fig.~\ref{qtb4}(a) shows the function $q_d(\tau)$ for $\beta~=~4$ and
several values of $\Gamma$. The correlation function in imaginary time
decreases monotonically as a function of $\tau$ in the interval
$[0,\beta/2]$. The initial slope increases with $\Gamma$. For large 
$\Gamma$ the decay is exponential, 
reflecting the presence of a well developed gap in the excitation
spectrum. Fig.~\ref{qtb4}(b) represents the $\Gamma$-dependence of the 
free-energy per spin for the same value of $\beta$. For $\Gamma \gg 
1$, 
it approaches a $T$-independent asymptote, $f=\Gamma/8$.

For $\beta > \beta_{\sc p}$, the solutions obtained for $\Gamma \gg 1$ and
$\Gamma \ll 1$ are still unique but they are not continuously connected.
The solution that derives from that for $\Gamma \gg 1$ ({\sc pm}$_1$)
can only be followed down to  a critical coupling  
$\Gamma = \Gamma_{c2}$ ,  where it ceases to
exist.  Conversely, the solution that exists for $\Gamma \ll 1$  ({\sc pm}$_2$)
can be followed only up to $\Gamma = \Gamma_{c1}$.
Since $\Gamma_{c2} <\Gamma_{c1}$ the two solutions coexist in the
interval $\Gamma_{c2} \le \Gamma \le \Gamma_{c1}$. Figure~\ref{qtb8}
shows the two types of solutions for $\beta = 8$ and several values
of $\Gamma$. For this temperature, the two
paramagnets coexist in the region  $2.8 \le \Gamma \le 3.6$. 
The corresponding free-energies 
are shown in Fig.~\ref{qtb8} (b). The free-energies cross at a point
intermediate between  $\Gamma_{c2}$ and  $\Gamma_{c1}$. Notice that, below the crossing
point,  the   classical {\sc pm} has the lowest free-energy. 
These results are reminiscent of those found within the 
static approximation with  {\sc pm}$_1$ and  {\sc pm}$_2$
corresponding to the solutions with $q_d = q_d^<$ and 
$q_d = q_d^>$ of Section \ref{static_approx}, respectively. 

\begin{figure}[ht]
\epsfxsize=3in
\centerline{\hbox
{
\epsffile{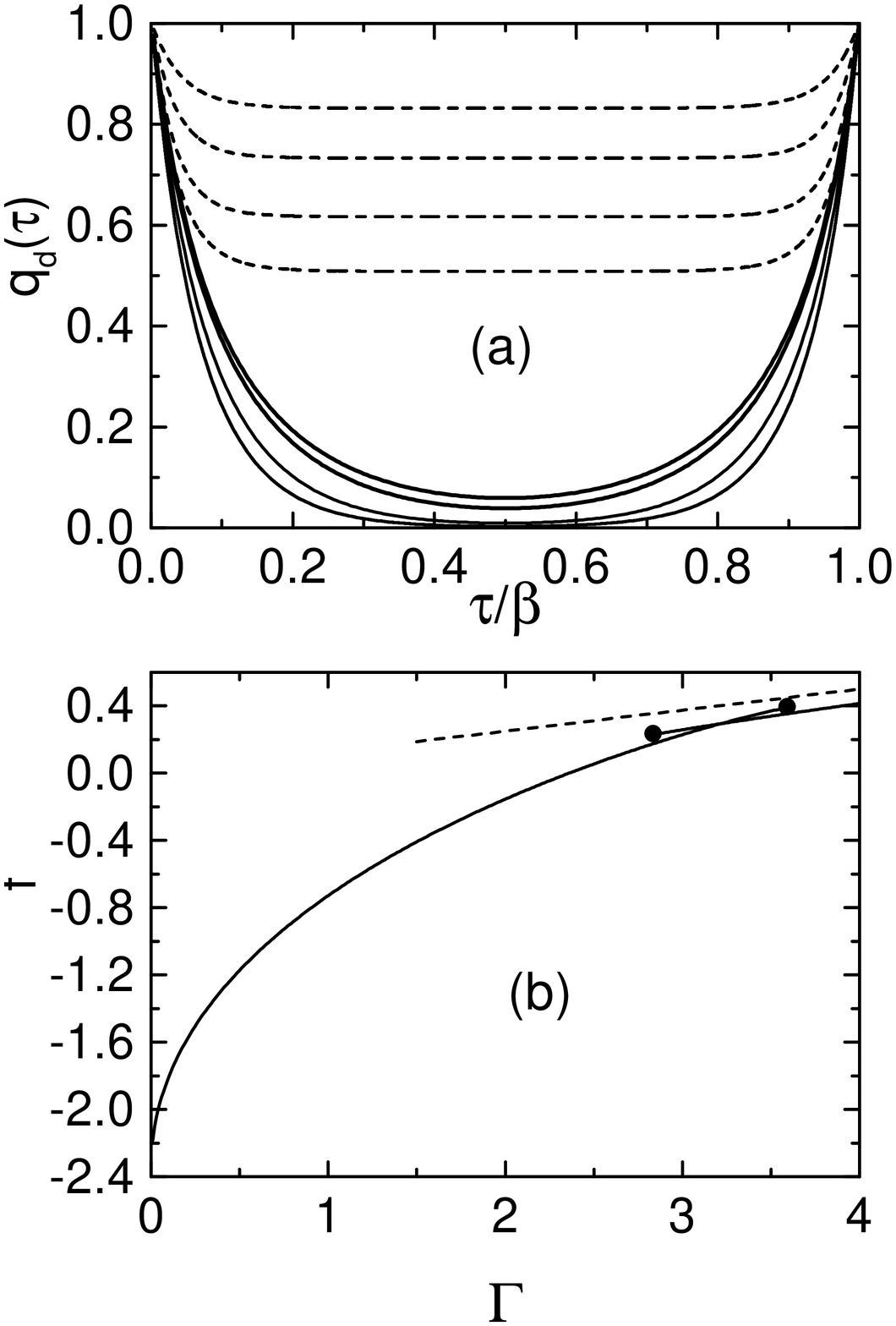}
}}
\caption{(a) $q_d(\tau)$ for the {\sc pm} solutions at
$\beta=8$. Solid lines: classical {\sc pm} solutions. $\Gamma=1, 2, 3, 3.5$
from top to bottom. Dashed lines: quantum  {\sc pm}
solutions. $\Gamma=2.8, 3, 3.5, 4$ from top to bottom.
(b) Solid Lines:  free-energies of {\sc pm} solutions as functions
of $\Gamma$ at $\beta=8$. The solid circles are the points at 
$\Gamma_{c1}=3.6$ and $\Gamma_{c2}=2.8$ where the  
solutions cease to exist. Dashed line: the asymptotic
value $f=\Gamma/8$.}
\label{qtb8}
\end{figure}

The two {\sc pm} states are qualitatively different. 
In the quantum {\sc pm}, 
$q_d(\tau)$ decreases
exponentially in $0 \le \tau \le \beta/2$. In the classical {\sc pm}, after a transient,
 $q_d(\tau)$ levels-off to a $\tau$-independent value
that decreases with increasing $\Gamma$. An analysis of the connection between
$q_d(\tau)$ and $\chi''(\omega)$ given in Eq.~(\ref{connection}) 
allows us to understand these differences in terms of the spectral properties
of the two states. We find that the exponential decay of the
correlation function of the quantum paramagnet reflects the presence
of a gap in $\chi''(\omega)$ and that the plateau that appears in the
case of the classical paramagnet reveals the
presence of a central peak with a narrow width  
$\Delta \omega \ll T$  in the gap. The hight of the plateau 
measures the fraction of the
total spectral weight under the peak.
 
\subsection{The spin-glass phase}

The equations that describe the {\sc sg} phase 
contain an extra parameter, the break
point $m$. 
The solutions are found as follows. 
Fixing $\beta$ and $m$, we compute $q_{\sc ea}$, 
${\tilde{q}}_d(0)$ and $z'$ from Eqs.~(\ref{eq:q1p}), (\ref{eq:qdp}) 
and (\ref{z'}), respectively. The value of $x_p$ is chosen according
to whether we want to study the equilibrium or the marginal {\sc sg}
states. We then solve Eq.~(\ref{chireg})
iteratively   varying 
$\Gamma$, starting from $\Gamma=0$,
 until a value is found that satisfies the spherical
constraint. 
For each temperature lower than the classical transition temperature,
 this procedure is repeated for different values of $m$.  
 
This procedure allows us to determine a function 
$m\equiv m(T,\Gamma)$. As shown in 
Fig.\,\ref{sdmgamma} for $\beta = 20$, $m(T,\Gamma)$ has two branches
 that meet at $\Gamma = \Gamma_{\rm max}(T)$. 
Physical values of $m$ lie on the  branch that  
satisfies that $m(T,\Gamma=0) = m_{\sc class}(T)$, the
classical
break point.
There are no solutions  of Eq.~(\ref{chireg}) for 
$\Gamma > \Gamma_{\rm max}(T)$ which may be identified as
 the value of the coupling above
which  
quantum fluctuations destroy the {\sc sg} phase. 

\begin{figure}
\centerline{\hbox
{
\epsfxsize=3in
\epsffile{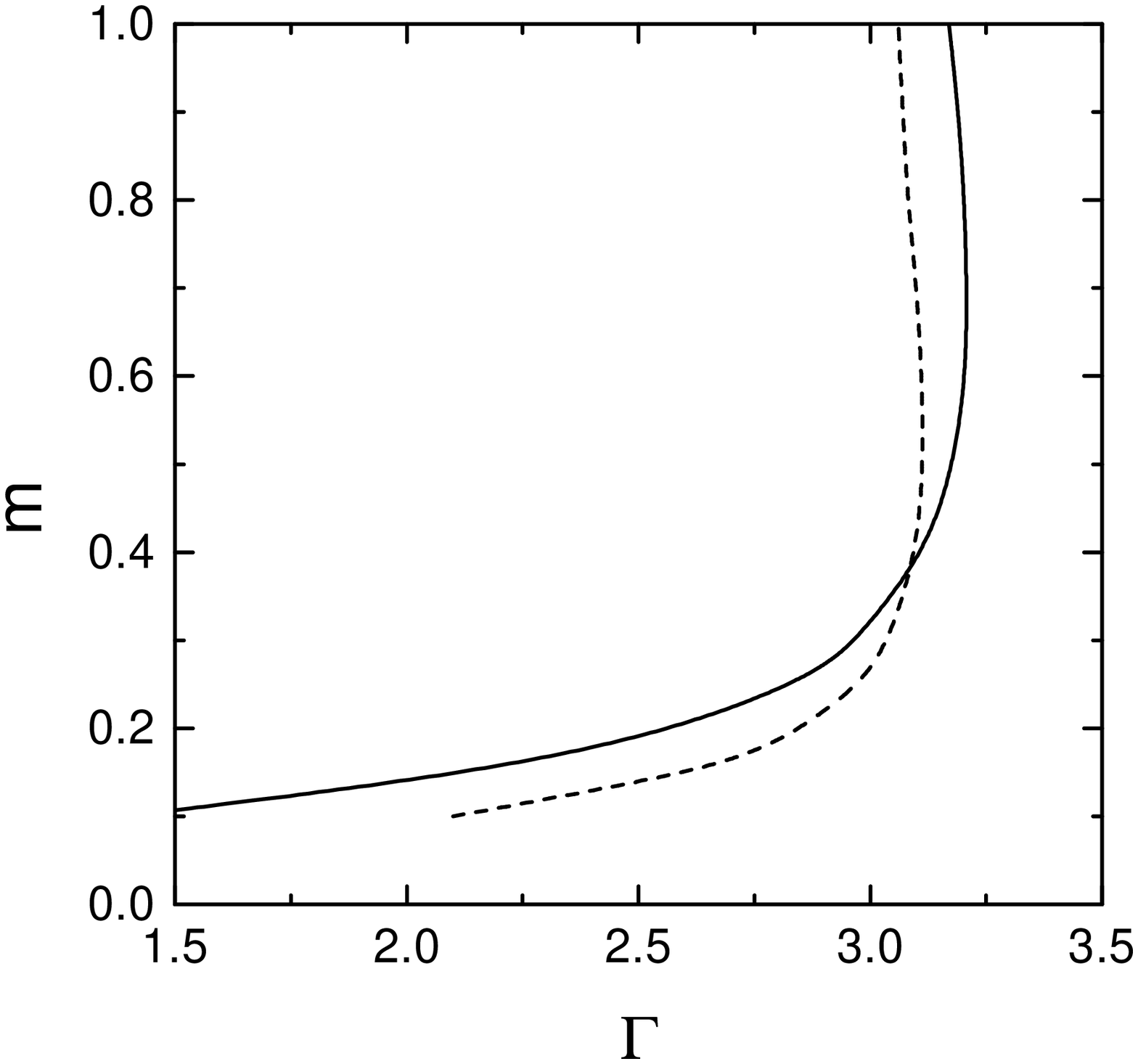}
}}
\caption{The breaking point 
$m$ as a function of $\Gamma$ for $p=3$ and $\beta =20$, 
both for the equilibrium (solid line) and marginally stable (dashed line) 
{\sc sg} state. For the equilibrium state $(\Gamma_{\sc max}, m_{\sc max}) 
\simeq (3.21, 0.7)$. For the marginally stable {\sc sg} state
$(\Gamma_{\sc max}, m_{\sc max}) \simeq (3.12, 0.5)$.}
\label{sdmgamma}
\end{figure}
We found that, as in the static approximation, there exists a
temperature $T^{\star}$ such that, for  $\beta \ge
\beta^{\star}$, $m_{\rm max}(T) <  1$. In all cases 
$q_{\sc ea}$ is finite at $\Gamma_{\rm max}(T)$. This fact and 
Eq.~(\ref{eq:q1p}) imply that $\lim_{T\to 0}  m_{\rm max}(T) =
0$, meaning    
that replica
symmetry is restored at the quantum critical point. The same feature 
was previously found 
in a study of  the SU(${\cal N}$) Heisenberg model~\cite{antoine}.

Fig.~\ref{figura4} shows the regular part of the auto-correlation
function (cf. Eq.~(\ref{def_reg})), for $\beta = 4$ and $\beta = 12$ and 
values of $\Gamma$ in the {\sc sg} phase. It may be seen that 
$q_{\sc reg}(\tau)$ decays more rapidly in the equilibrium state than
in the marginally stable state. Analysis of the curves of
Fig.~\ref{figura4} for $\beta = 12$
shows that, for $1 \ll \tau
\ll \beta/2$, $q_{\sc reg}(\tau)$ decays exponentially in the first case
but $q_{\sc reg}(\tau) \propto \tau^{-2}$ in the second one.  This
follows again from the differences in  the excitation spectra of
the two states. 

\begin{figure}
\centerline{\hbox
{
\epsfxsize=3in
\epsffile{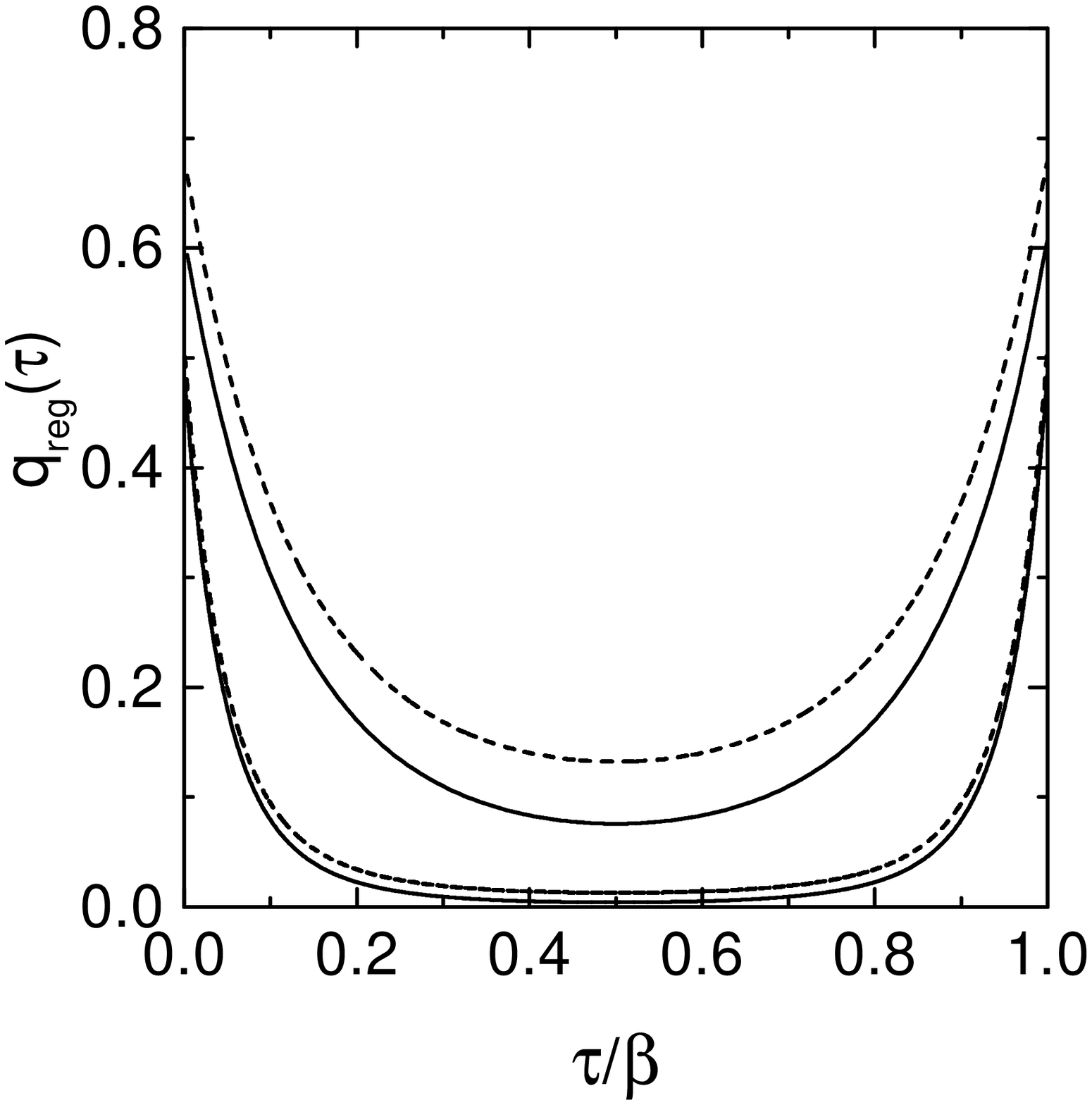}
}}
\caption{Regular part of the auto-correlation function in the {\sc sg}
phase. Full and dashed lines represent the equilibrium and marginally
stable states, respectively. Upper curves: $\beta = 4$, $\Gamma =
2.35$.
Lower curves: $\beta = 12$, $\Gamma = 1.75$. }
\label{figura4}
\end{figure}

In the case of the marginally stable solution, 
$q_{\sc reg}(\tau)$ may also be computed using the low-frequency
approximation of Section~\ref{approx_sol}. 
The exact and approximate
 results for $\beta=20$ are shown in Fig.~\ref{figura5}. It can be seen that the
agreement between the two 
is very good.

\begin{figure}
\begin{center}
\epsfxsize=3in
\epsffile{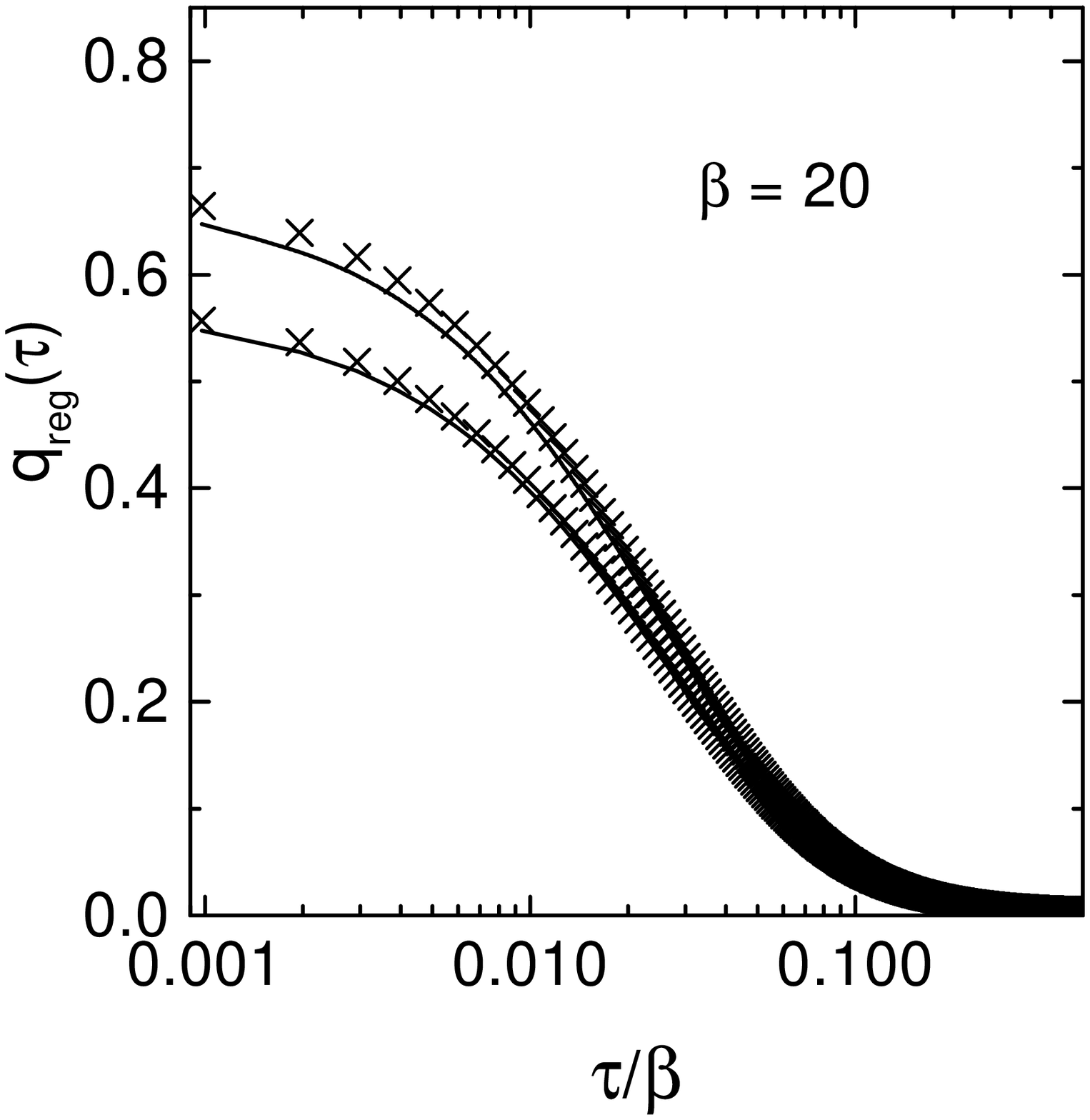}
\end{center}
\caption{Comparison of the exact numerical solution for 
$q_{\sc reg}(\tau)$ (solid lines) and the approximation of 
Section~\ref{approx_sol}
(symbols) for $\beta = 20$. Upper curves: $\Gamma = 2.0$. Lower curves
$\Gamma = 3.0$.}
\label{figura5}
\end{figure}

\subsection{Equilibrium phase diagram}
\label{ph_di}

In this section we discuss the phase diagram that results from the
analysis of the equilibrium solutions. The case of the
marginally stable solutions will be discussed in the next Section. 
As within the static approximation, all the {\sc pm} states that we
found are locally stable. Their free-energies  must be compared with 
that of the {\sc sg} state in order to
construct the equilibrium phase diagram. As before, 
there exists a temperature interval $T^{\star} \le T \le T_{\sc p}$
within which there are two first order transitions as a function of
$\Gamma$, one from the quantum to the classical {\sc pm}, followed at
a lower coupling by the {\sc sg} transition. However, for $p=3$, the
case for which we have performed detailed numerical calculations, $T_{\sc p}
- T^{\star}$ is very small and, in the following, we only consider in detail
the cases $T > T_{\sc p}$ and $T < T^{\star}$. 
Figures~\ref{free-energies}(a) and (b) show the $\Gamma$-dependence of
the {\sc pm} and {\sc sg} free-energies for $p=3$ computed using 
Eq.~(\ref{eq:fe1}) for two temperatures,
above $T_{\sc p}$ and below $T^{\star}$. Solid lines and symbols
represent the {\sc pm} and {\sc sg} solutions, respectively.  The curves end at the point
where the corresponding solution disappears.
It may be seen that for $T>T_{\sc p}$  the free-energies of the two
states merge precisely at $\Gamma_{c}(T)$ where $m=1$. 
Below the critical point,
$f_{\sc sg} > f_{\sc pm}$ meaning that the {\sc sg} solution
{\it maximizes} the free-energy. As in the static approximation, 
stability arguments do not exclude 
the {\sc pm} solution below $\Gamma_c$ as a metastable state but we
may 
resort to the arguments given in Section
\ref{static_approx} to argue that the continuation of the {\sc pm}
solution into the  {\sc sg} phase is unphysical also in this
case.
\begin{figure}
\centerline{\hbox{
\epsfxsize=3in
\epsffile{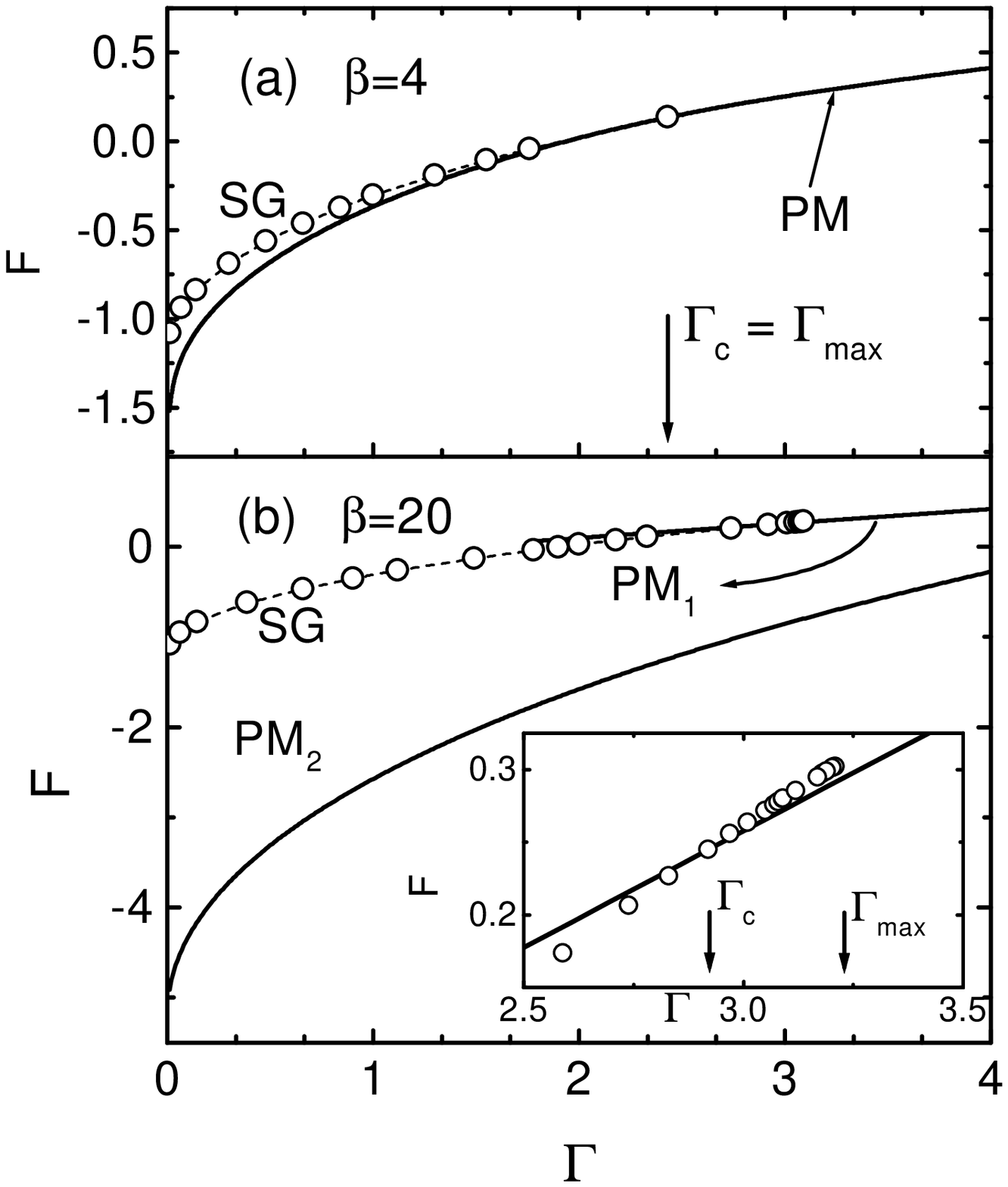}
}}
\caption{Free-energies of the different {\sc pm} (solid lines) and SG
(symbols) phases above $T_{\sc p}$ (a)
and below $T^{\star}$ (b). The inset in panel (b) shows in detail the
crossing of the free-energies at the critical point for $T < T^{\star}$.}
\label{free-energies}
\vspace{1cm}
\centerline{\hbox{
\epsfxsize=3in
\epsffile{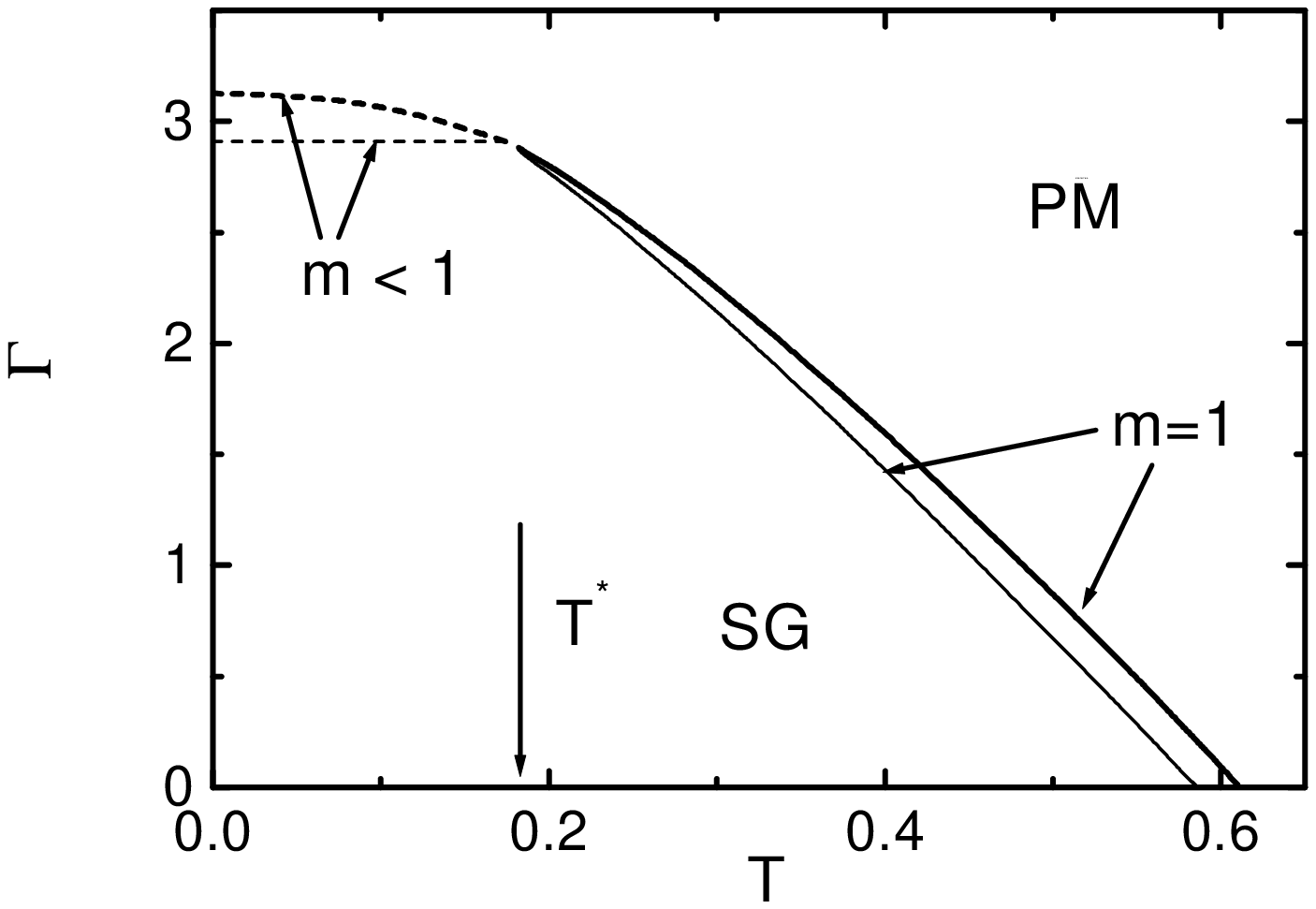}
}}
\caption{Static (thin lines) and dynamic (thick lines) phase diagrams
of the $p$-spin model for $p=3$. Solid and dashed lines represent
second and first
order transitions, respectively.}
\label{phase}
\end{figure}

Below $T^{\star}$ we must compare the free-energies of the two {\sc pm} 
solutions with that of the {\sc sg} state to choose the most favorable
phase. In this temperature range, the 
classical paramagnet has the same unphysical
properties that we found in Section \ref{static_approx} in the static
approximation: the ground-state energy and susceptibility diverge and its free
energy does not intersect that of the ordered state. 
Therefore, we 
discard it in favor of {\sc pm}$_1$ even if its free-energy is
higher. 
The free-energies of the {\sc sg} and {\sc pm}$_1$ states cross at 
$\Gamma_{c}~<~\Gamma_{\rm max}$ as shown in the inset in Fig.~\ref{free-energies}(b).
 As in the static approximation, 
${f}_{\sc sg} < {f}_{\sc pm}$ for  $T < T^{\star}$
whereas ${f}_{\sc sg} > {f}_{\sc pm}$ for 
$T > T^{\star}$.
The {\sc sg} and {\sc pm}$_1$ solutions extend beyond the point
where their free-energies cross and there is phase coexistence. 
Now, both $q_{\sc ea}$ and $m$ are
discontinuous at $\Gamma_{c}$. The {\sc sg} transition is thus {\it
first} order with latent heat and discontinuous
susceptibility (see below). The phase diagram resulting from this analysis is
represented in Fig.\,\ref{phase} (thin lines). The flat
section is the first-order line, which {\it does not} exhibit the 
spurious reentrant behavior found within the static approximation  
(Fig.~\ref{fig:phasedia}). In addition, in contrast with the approximate
phase diagram, the transition line has zero-slope at $T=0$, which is a
consequence of the third law of thermodynamics as shown in Ref.~\cite{Niri}.

We also computed the $\Gamma$-dependence of  $q_{\sc ea}$
and the static susceptibility,
\begin{equation}
\chi=\int_0^{\beta} d\tau \; [q_d(\tau) - (1-m) q_{\rm
EA}]
\; , 
\end{equation}
as functions of $\Gamma$ for the  $p=3$ model.
The results are displayed in Fig.\,\ref{chi}.
The susceptibility has a cusp at $\Gamma_{c}$ for $T > T^{\star}$
and a discontinuity
for $T < T^{\star}$. The dotted lines correspond to the regions of
metastability. Notice that the susceptibility is {\it higher} on the
{\sc sg} side of the transition, an unusual result. This follows from 
 the  fact that the gap in the excitation spectrum of the
{\sc pm} state is wider than that of the {\sc sg} state. For the same
reason the entropy of the {\sc pm} is {\it lower} leading to a {\it
negative} latent heat at the
transition.

As shown in the lower pannel of  Fig.\,\ref{chi}, 
the order parameter decreases rapidly with increasing $\Gamma$ and is 
reduced by a factor of two half-way from the
transition, showing that quantum fluctuations are quite strong in this
system. 

\begin{figure}
\epsfxsize=3in
\centerline{\hbox{
\epsffile{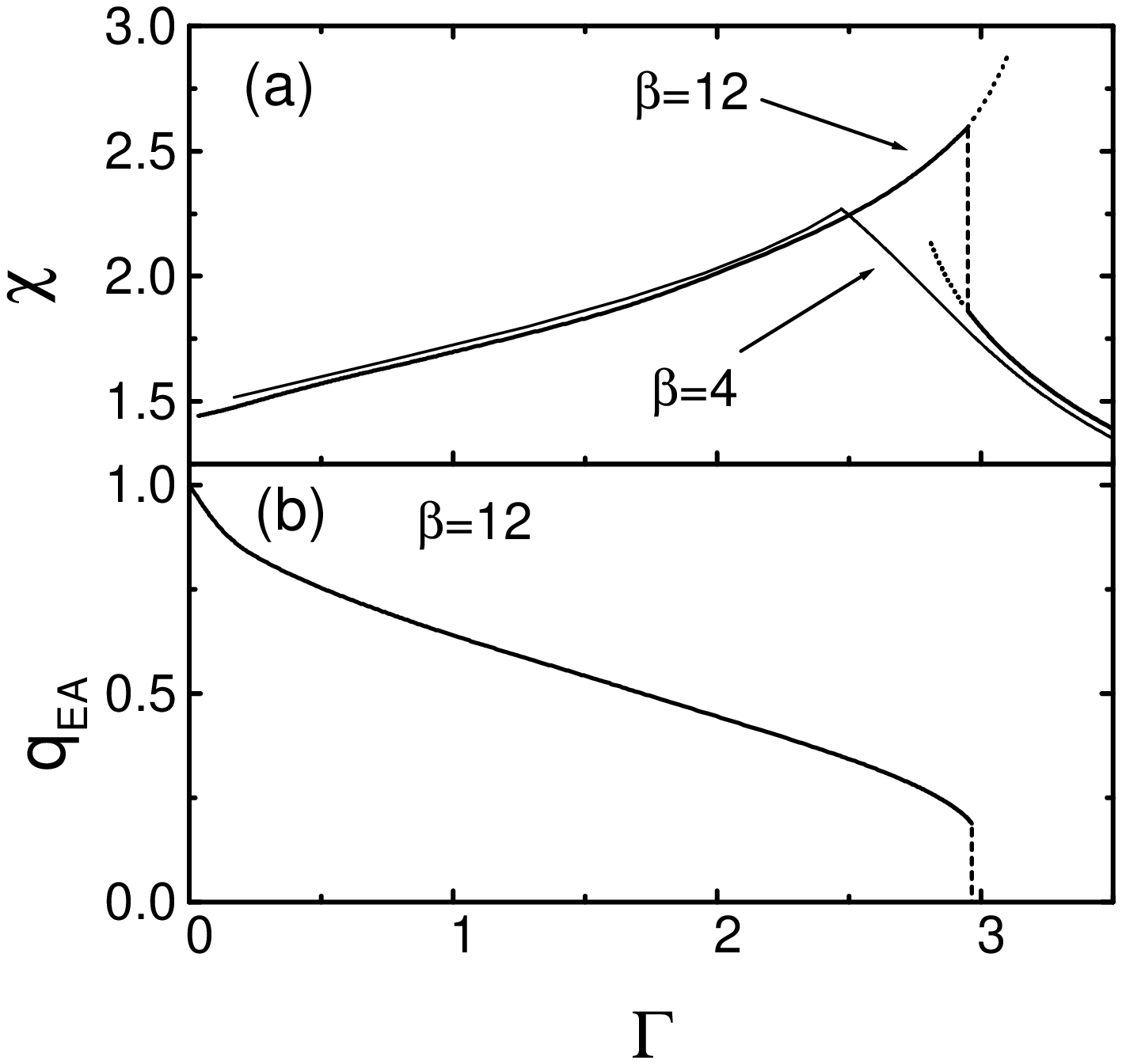}
}}
\caption{Magnetic susceptibility (a) and Edwards-Anderson order
parameter (b) of the $p$=3 model.}
\label{chi}
\end{figure}

As mentioned above, early experiments \cite{aeppli1} 
on LiHo$_x$Y$_{1-x}$F$_4$ in a transverse magnetic field 
suggested that the second
order {\sc sg} transition observed for moderate fields could become
first order for $H \approx H_c(0)$, the field above which the
system remains  {\sc pm} down to zero temperature.  
The recent observation of hysteresis effects in the
transverse field-dependence of the susceptibility give 
 further support to this idea~\cite{aeppli2}.
While the model that we study here is not intended to describe
microscopically this compound, it captures some of its phenomenology. 
However, important differences between the
predictions 
of this model and experiment exist, the most important ones being the 
fact that, in the model, where the transition is second
order, it {\it is not} accompanied by a divergence of $\chi_3$ and
that
 the jump 
of $\chi$ across the first order transition line has a sign 
opposite to that observed experimentally.

\section{Connections between the marginally 
stable state and real time dynamics}
\label{real_time}
\setcounter{equation}{0}
\renewcommand{\theequation}{\thesection.\arabic{equation}}

As is well known, the dynamics of {\it classical} spin-glasses 
becomes non-stationary below a dynamic  
transition temperature $T_d$  that may or may not coincide with
$T_g$. 
In the low-temperature phase, 
the systems shows {\it aging}, {\it i.e.}, the correlation function $C(t,t')$ depends
both on the time-difference $(t - t')$ and on $t'$~\cite{Fihu,Bo,Cuku}. 
Aging has experimental manifestations such as a dependence of
the ac-susceptibility on frequency and the time elapsed after a
quench. Theoretical studies of aging effects in quantum
systems have recently appeared in the 
literature~\cite{leti1,Mapo,Igri,Chamon}. Some systems where 
quantum fluctuations might play a role, like magnetic 
nanoparticles~\cite{nanoparticles} and disordered $2D$-electronic
systems \cite{Zvi}, were investigated experimentally. 
A search for these effects in the Li(Ho/Y)F$_4$
system is under way  \cite{aeppli2}.

The existence of connections between dynamic and static calculations 
in classical disordered
systems is known since
the work of Sompolinsky  on the equilibrium dynamics of
the SK model \cite{So}. Apart from some conceptual problems 
that have been extensively discussed in 
the literature~\cite{Hojayo}, Sompolinsky noted that his dynamic solution 
for the classical SK model was marginally stable. 
Kirkpatrick and Thirumalai  \cite{KT} showed that, in the classical 
mean-field $p$-spin-glass model, $T_d$ can be calculated in
 a fully static replica approach if the break point $m$ is computed by 
 demanding that the spin-glass susceptibility 
$\chi_{\sc sg}$ diverges, instead of asking that the free-energy be an
extremum with respect to variations of this parameter. In technical terms, 
this prescription corresponds to setting the replicon eigenvalue
to zero~\cite{Bray-Moore-replicon} and goes under the name of 
{\it condition of marginal stability}. 
It has already been used within 
the Matsubara approach to quantum disordered problems 
on the basis of the physical assumptions of having a gapless 
solution~\cite{ledou,antoine}.

In this Section we compare the results obtained above using the marginal stability
criterion with the outcome of a  full dynamical calculation of the
same system {\it weakly} coupled 
to a quantum bath made of an ensemble of harmonic oscillators \cite{leti1}. 
The Hamiltonian of the coupled system is 
\begin{equation}
H = H_{\sc syst} + H_{\sc bath} + \alpha H_{\sc int}
\; ,
\end{equation}
with $H_{\sc syst}$ given in (\ref{eq:action}).
$H_{\sc bath}$ is the Hamiltonian of an ensemble of 
quantum harmonic oscillators with an
Ohmic spectral density. $H_{\sc int}$ represents a linear coupling between the 
spins and the oscillators' displacements and it is controlled
by a  
coupling constant $\alpha$. The oscillators are in equilibrium at
 temperature $T$ at all times and act as a reservoir.

The real-time evolution of the
system was derived using the Schwinger-Keldysh formalism and it is 
described by coupled differential equations for $C(t,t')$ and
$R(t,t')$, the symmetrized correlation and response
functions of the system, respectively \cite{leti1}. In the glassy phase, $C$
exhibits a two-time dependence of the form:
\begin{equation}
C(t+t_w,t_w) = C_{\sc st}(t) + C_{\sc ag}(t+t_w,t_w)
\label{eq:corr}
\; .
\end{equation}
The first term gives the time-dependent evolution for time-differences
 $t\ll t_w$ and varies between $1-q_{\sc ea}$ to $0$.
In this time regime, the second term is constant and equal to
  $q_{\sc ea}$ but decays 
from $q_{\sc ea}$ to $0$ in a waiting-time dependent manner when $t \gg 
 t_w$.    
More formally, the Edwards-Anderson parameter is defined as 
\begin{equation}
q_{\sc ea} \equiv \lim_{t\to\infty} \lim_{t_w\to\infty} C(t+t_w,t_w) 
\; .
\end{equation}
The response function can also be written by separating two terms as in (\ref{eq:corr})
\begin{equation}
R(t+t_w,t_w) = R_{\sc st}(t) + R_{\sc ag}(t+t_w,t_w)
\; .
\label{eq:resp}
\end{equation}
The first term is invariant under time-translations and it is related to the 
first term in (\ref{eq:corr}) by the quantum fluctuation-dissipation
theorem (FDT). The second
term is related to the second term in (\ref{eq:corr})
by a modified quantum FDT relation where the temperature of the bath $T$ is replaced by 
an effective temperature $T_{\sc eff} \equiv T/X$ which defines $X$,
the FDT violation factor \cite{Cukupe}.  In the long-$t_w$ limit,
\begin{equation}
R_{\sc ag}(t+t_w,t_w) = \frac{X}{T} \left. 
\frac{\partial C_{\sc ag}(t+t_w,t')}{\partial t'} \right|_{t'=t_w}
\; .
\end{equation}
For the model under consideration, weakly coupled to a bath, 
the FDT violation factor and $q_{\sc ea}$ are given by \cite{leti1}

\begin{eqnarray}
\beta X &=& \frac{p-2}{q_{\sc ea}} \, R_{\sc st}(\omega=0)
\equiv 
 (p-2)\sqrt{{2\over p(p-1)}}
q_{\sc ea}^{-p/2}
\; ,
\label{eq:teff_dyn}
\\
1 &=& \frac{p(p-2)}{2} q_{\sc ea}^{p-2} \, R^2_{\sc st}(\omega=0)
\; .
\label{eq:qea_dyn}
\end{eqnarray}
These equations have the same form as Eqs.~(\ref{eq:rep2}) and
(\ref{eq:rep3}) with $m$  and $\tilde{q}_{\sc reg}(0)$ replaced by $X$
 and  $R_{\sc st}(\omega=0)$, respectively.
It will be shown below that, in the limit $\alpha \to 0$,  the
equation obeyed by $R_{\sc st}(\omega)$ reduces to the analytic
continuation of Eq.~(\ref{chireg}). Therefore,  in this limit, 
$R_{\sc st}(\omega=0) = \tilde{q}_{\sc
reg}(i\omega_k=0)$. It follows from this fact and 
Eqs.~(\ref{eq:rep2})-(\ref{eq:rep3}) and 
(\ref{eq:teff_dyn})-(\ref{eq:qea_dyn}) that $m = X$. Although
the coincidence between the values of $X$ and $m$ for the  marginal
{\sc sg} 
state has been 
noticed several times for classical models, this is the first explicit
evidence 
of its validity in a quantum problem. 

This correspondence allows us to identify the  dynamic 
transition line $\Gamma_d(T)$ as the boundary in the $T-\Gamma$ plane
where the marginally stable solution exists. $\Gamma_d(T)$ is a piecewise curve 
formed by a line on which $X=1$ that joins $(T_d,0)$ and
$(T^{\star},\Gamma^{\star})$ and a line on which $X<1$ that joins 
$(T^{\star},\Gamma^{\star})$ and $(0,\Gamma_c)$.
The dynamic phase diagram for $p=3$ is
shown in Fig.\,\ref{phase} (thick lines). It is qualitatively similar
to that obtained in the static approximation. In particular, $m$ is
discontinuous across the dashed line and $\Gamma_d$ lies always
above $\Gamma_{c}$. Notice that the two transition lines come
very  
close to one another for $T \sim T^{\star}$ and seem to 
touch at $T^{\star}$. For $T < T^{\star}$, $m$ varies continuously along
$\Gamma_{\rm d}(T)$ and vanishes at the quantum
critical point\cite{footnote2}. 
This has a consequence of potential interest for
experiment: at low temperatures, FDT violations are predicted to appear suddenly as
 $\Gamma_{\rm d}$ is crossed coming
from the high $\Gamma$ region.

The equivalence between the replica theory supplemented 
with the marginality
condition ({\sc rtmc}) and the  dynamical approach goes beyond
the equality of $m$ and $X$. It also holds for  
the expectation values of operators calculated within {\sc rtmc} and 
 the long-time limit of the same observables in the coupled 
dynamical system, for $\alpha \to 0$. 
It must be stressed that the weak-coupling limit must be taken {\it after} the
thermodynamic
 and long-time limits \cite{Vieira}. More precisely, one can establish 
that for an operator $O$ depending only on the coordinates and momenta
of the system, 
\begin{equation}
\langle O \rangle_{\sc rtmc} 
=
\lim_{\alpha\to 0} \lim_{t\to\infty} \lim_{N\to \infty} 
\langle O(t) \rangle_{\sc dyn}
\; ,
\end{equation}
where 
$\langle \;\cdots\;\rangle_{\sc rtmc}$ is the outcome of the {\sc rtmc}
calculation.

To illustrate this point, we consider Eq.~(\ref{chireg}) fixing 
$p=3$ for simplicity. We define the real-time quantities
\begin{eqnarray}
\label{continuation}
\displaystyle{
\begin{array}{rcl}
\tilde{R}_{\sc rtmc}(\omega)=\left. \tilde{q}_{\sc reg}(i
\omega_k)\right|_{i \omega_k \to \omega + i 0^+} \;, & R_{\sc rtmc}(t) =  
\int_{-\infty}^\infty \frac{d\omega}{2\pi}\;\tilde{R}_{\sc
rtmc}(\omega)\exp(-i \omega t)\; , 
\\
\\
\tilde{\Sigma}_{\sc rtmc}(\omega)=\left. \tilde{\Sigma}_{\sc reg}(i
\omega_k)\right|_{i \omega_k \to \omega + i 0^+} \;, & 
\Sigma_{\sc rtmc}(t) =  
\int_{-\infty}^\infty \frac{d\omega}{2\pi}\;\tilde{\Sigma}_{\sc
rtmc}(\omega)\exp(-i \omega t)\; , 
\end{array}
}
\end{eqnarray}
and introduce the Lehmann representations
\begin{eqnarray}
\label{lehmann} 
\tilde{q}_{\sc reg}(i \omega_k) = \int_{-\infty}^\infty \frac{d\omega}{
\pi}\frac{{\rm Im}\; \tilde{R}_{\sc rtmc}(\omega) }{\omega -
i\omega_k} \;,
\;\;\;\;\;\;\;
 \tilde{\Sigma}_{\sc reg}(i \omega_k) = \int_{-\infty}^\infty \frac{d\omega}{
\pi}\frac{{\rm Im} \; \tilde{\Sigma}_{\sc rtmc}(\omega) }{\omega - i\omega_k} \; .
\end{eqnarray}
It follows from Eqs.~(\ref{continuation}) and (\ref{lehmann}) that
\begin{eqnarray}
\label{intime} 
\begin{array}{rcl}
R_{\sc rtmc}(t) = \theta(t) \int_{-\infty}^\infty \frac{d\omega}{
\pi} \; {\rm Im}\, \tilde{R}_{\sc rtmc}(\omega) \sin(\omega t)\; ,
\\
\\
\Sigma_{\sc reg}(t) =  \theta(t) \int_{-\infty}^\infty \frac{d\omega}{
\pi} \; {\rm Im} \, \tilde{\Sigma}_{\sc rtmc}(\omega) \sin(\omega t)\; .
\end{array}
\end{eqnarray}
Analytically continuing Eq.~(\ref{chireg}) and Fourier transforming
the resulting equation we arrive at
\begin{equation}
\label{real_time_response}
\left[\frac{1}{\Gamma}\;\frac{\partial^2}{\partial t^2} + z_{\sc rtmc}
\right] R_{\sc rtmc}(t) = \delta(t) + \int_{0}^t dt'\;\Sigma_{\sc
rtmc}(t - t')  R_{\sc rtmc}(t')\; ,
\end{equation}
where the limits on the integral on the right-hand side follow from
Eq.~(\ref{intime}) 
 and 
\begin{equation}
\label{newz}
z_{\sc rtmc} = z' + \tilde{\Sigma}_{\sc reg (0)} \equiv  R_{\sc rtmc}^{-1}(\omega = 0)
+  \int_{0}^{\infty} dt'\;\Sigma_{\sc
rtmc}(t').
\end{equation} 
The last equality follows from the combined use of Eqs.~(\ref{z'}),
Eqs.~(\ref{eq:q1p}) and ~(\ref{handyequation}).
Rewriting Eq.~(\ref{eq:diff1}) for $p=3$ in the form 
\begin{equation}
\tilde{\Sigma}_{\sc reg}(i
\omega_k) = \frac{3}{2 \beta} \sum_{\omega_m} \tilde{q}_{\sc
reg}(i \omega_m)\;\tilde{q}_{\sc reg}(i \omega_k - i \omega_m) + 3
q_{\sc ea}\;\tilde{q}_{\sc
reg}(i \omega_k) \; ,
\end{equation}
and using Eq.~(\ref{lehmann}), we find after some algebra 
\begin{equation}
\label{imsigma}
{\rm Im} \; \Sigma_{\sc rtmc}(\omega) = \frac{3}{2}
\int_{-\infty}^{\infty} 
\frac{d\omega'}{\pi} C_{\sc rtmc}(\omega')\; 
{\rm Im}\; R_{\sc rtmc}(\omega' - \omega)
\; ,
\end{equation}
where we have introduced the Fourier transform of the symmetrized
correlation function
\begin{equation}
\label{scf}
C_{\sc rtmc}(\omega) =  {\rm
Im}\; R_{\sc rtmc}(\omega)\; \coth\frac{\beta \omega}{2} +
2\pi\;q_{\sc ea}\;\delta(\omega).
\end{equation}
It follows that
\begin{equation}
\label{selfenergy}
\Sigma_{\sc rtmc}(t) = 3\;\theta(t) C_{\sc rtmc}(t) \;R_{\sc rtmc}(t)
\equiv - 3\;{\rm Im} \left[C_{\sc rtmc}(t) - \frac{i}{2} R_{\sc
rtmc}(t)\right]^2\; .
\end{equation}
Equations~(\ref{real_time_response}), (\ref{newz}) and  (\ref{selfenergy}) are equivalent to 
Eqs.~(6.6), (6.16), (7.13) and (3.27) of Ref.~\cite{leti1} specialized
to $p=3$ and taken in the limit $\alpha \to 0$. It follows that 
\begin{equation}
\label{compar}
R_{\sc rtmc}(t) = \lim_{\alpha\to 0} \lim_{t_w\to\infty} R_{\sc dyn}(t
+ t_w,t_w) 
= \lim_{\alpha\to 0} R_{\sc st}(t) \; .
\end{equation}
The equality of the {\sc rtmc} response function and the stationary
part of the dynamical response function implies that all the quantities
that can be expressed in terms of them are also equal. In particular, it can be shown
that the asymptotic dynamic energy-density of the system in contact with
the bath in the weak coupling limit is precisely given by Eq.~(\ref{internal_sg}).

\section{Conclusions and discussion}
\setcounter{equation}{0}
\renewcommand{\theequation}{\thesection.\arabic{equation}}
\label{conclusions}

In this paper we presented a detailed study of the properties of 
the quantum spherical $p$-spin-glass model in thermodynamic
equilibrium and in the marginally stable state.

We solved the equations that describe the different phases of the
 system numerically and also using various approximation schemes that
 give  valuable physical insights, in particular, on
the delicate issue of handling the multiplicity of
solutions of the equations.  

We established explicitly a connection between the states obtained in 
the quantum replica calculation with the marginality 
condition and those that result form the  non-equilibrium dynamics of
the system  weakly coupled to an  environment 
in thermal equilibrium.

We determined the phase diagram of the model and showed that, in all 
cases, quantum fluctuations drive the {\sc sg}
transition first-order at sufficiently low temperatures. 
A tricritical point separates the two 
transition lines. This shows that the
properties of {\sc sg} systems in the quantum regime can be
qualitatively different from those in the classical limit.
The same feature has been first observed in the $p\to\infty$ limit of 
the $p$-spin model in a transverse field~\cite{Gold} and in a 
quantum model with multiple 
interactions studied in Ref.~\cite{Niri}. These studies, as well as 
 the TAP analysis of Ref.~\cite{Bicu}, suggest that this phenomenon   
may be rather generic in quantum extensions of models having a 
{\it discontinuous} transition in the classical limit.

We notice, however,
 that Ritort studied the phase transition in  the 
random orthogonal model (ROM)~\cite{ROM} in a transverse field 
within the static approximation and found no evidence
of the existence of a tricritical point~\cite{Ritort1}.
In this case, an  expansion of the 
free-energy in powers of $m-1$ leads to a vanishing 
Edwards-Anderson parameter at the quantum critical point
$(0,\Gamma_c)$~\cite{Ritort2}. 
It would be interesting to investigate whether this 
behavior subsists beyond the static approximation, {\it i.e.} 
whether a tricritical
point at finite temperature appears in the exact solution. 

A quantum particle (or a manifold) in an infinite dimensional 
random potential has a discontinuous classical transition when the 
potential has short-range correlations~\cite{random-manifold}. 
Therefore, the quantum extensions analyzed by Goldschmidt~\cite{Gold2} and 
Giamarchi and Le Doussal~\cite{ledou} 
are hence candidates to exhibit a first order transition 
in the limit of zero temperature. 

So far, studies of quantized models with {\it continuous} 
classical transitions have found continuous transitions
  close to the quantum critical point. However,
we believe that this issue should be explored further. Full
solutions of the low-temperature phase of these models do not yet
exist. They would  
need a full replica symmetry breaking {\it Ansatz} for the non-diagonal 
terms in the replica matrix. In our opinion the soft spin quantum
model whose real-time dynamics has been studied by Kennett and 
Chamon~\cite{Chamon} is the simplest choice to 
investigate this question.

Many of the results of this paper can also be obtained 
using the TAP approach~\cite{Bicu}.
This formalism, that avoids the use of replicas, 
allows us to understand the 
organization of the metastable states and to interpret the dynamic
behavior of glassy systems in terms of them.  
 The relationship
between the quantum TAP equations and the Matsubara approach for this
model was analyzed In Ref.~\cite{Bicu}. A 
relationship between $(q_{\sc eq},m,q_d(\tau))$ in the replica approach
and $(q_{\sc eq},{\cal E},C(\tau))$ in the TAP approach was obtained
where 
${\cal E}$
is the energy density of a TAP solution and $C(\tau)$ the
imaginary-time dependent correlation function. Different choices for
$m$, such as the equilibrium condition the marginality condition or
others, 
correspond to different choices for ${\cal E}$: the equilibrium value,
the value reached dynamically or some intermediate one, respectively. 

In Section~\ref{comparison} we mentioned that this model can be 
viewed as a description of a  closed polymer embedded in 
an infinite dimensional 
space with a random potential~\cite{Ha-he,Baum,Gold3}. 
The line tension favors a linear
configuration of the polymer while the disorder makes the 
polymer wander and search for configurations that are favorable 
from an energetic point of view. As seen above, besides a transition 
line separating a liquid-like phase from a 
glassy-like phase, we expect a transition line between two different 
liquid phases corresponding to the two paramagnetic 
solutions that we found in the spin-glass model. 
One of the liquid phases is characterized by 
a large value of the correlation of two monomers separated at 
maximum distance $L/2$, with $L$ the total length of the polymer. 
(In the static approximation
we called this solution $q_d^>$.)  In the polymer 
language, the other solution corresponds to a state in which 
the correlation 
of two monomers at distance $L/2$  is small and close to zero at
high values of $\Gamma$. These two configurations can be interpreted 
as representing a coiled polymer ($q_d^>$) and a linear-like polymer ($q_d^<$). 
These results seem reasonable since the solution $q_d^<$ appears at
high values of $\Gamma$, {\it i.e.} when the polymer is very flexible, while 
the solution $q_d^>$ appears at low values of $\Gamma$ when the
polymer is more rigid. The nature of the transition from the 
liquid-like to the glassy-like  phase is different in the two
situations. This transition is expected to occur on a sizable 
region of phase space only for $p \gg
1$ which corresponds to short range correlations of the random
potential.
  It would be interesting to check whether such a 
crossover occurs in finite dimensional models of  random 
directed polymers and in experiments.

This work can be extended in several directions. 
One of them is the question of how the choice of particular
 initial conditions affect the real-time dynamics of these disordered 
quantum models. Up to now, only the case of random initial conditions,
which simulate an instantaneous quench from the high-temperature phase,
was discussed in the literature~\cite{leti1}. An interesting problem
is  
the relaxation dynamics within metastable states that are 
solutions of the quantum {\sc TAP} equations \cite{Bicu}, in the manner
of Refs.~\cite{Hojayo} and \cite{barrat}. To solve this
problem   
one has to represent the generating functional of the correlation
functions in the form of a path integral 
on a time  contour that mixes real and imaginary times. 
The derivation of the relevant equations and their solution are  not 
straightforward and  need, as input, the solution of the imaginary-time
equilibrium equations studied in this paper. We shall discuss this
problem
 elsewhere \cite{LDCass}.
Another open problem is how the coupling to a strong quantum environment
modifies the statics and dynamics of such disordered models, an issue
that we shall discuss  in a separate publication \cite{new}.

\section{Acknowledgements}
Useful discussions
 with G. Biroli,  T. Garel, A. Georges,  J. Kurchan,
L. B. Ioffe, G. Lozano, D. Mukamel, O. Parcollet, F. Ritort,
M. Rozenberg and
V. R. Vieira are gratefully acknowledged.  We thank particularly
 G. Aeppli and T. Rosenbaum for making their results
available to us prior to publication. LFC and DRG thank the
 ECOS-Sud programme for
a travel grant. LFC acknowledges financial support from the ACI
``Algorithmes d'optimisation et syst{\`e}mes d{\'e}sordonn{\'e}s
quantiques''. CAdSS is financially supported by the
Portuguese Research Council, FCT, under grant PRAXIS XXI/BPD/16303/98.  

\section *{Appendix}
\setcounter{equation}{0}
\renewcommand{\theequation}{A.\arabic{equation}}
\label{appendix}

In this appendix, we demonstrate that, 
for the model studied in this paper, the extremization equations for a 
2-step {\sc rsb} {\it Ansatz} collapse into the ones for the 1-step 
{\sc rsb}. This procedure can be iterated for a $k$-step 
{\sc rsb} in order to prove that  the 1-step {\sc rsb} scheme is 
exact.

Within a $k$-step {\sc rsb} {\it Ansatz}, the $n$ replicas are organized 
into $n/m_1$ ``1-families'' of $m_1$ elements each. The $m_1$ 
replicas in a ``1-family'' are then further organized into $m_1/m_2$ 
``2-families'' of $m_2$ elements each, and so on until the level 
of ``$k$-families'' is reached. We use the convention that all 
replicas belong to the same ``0-family'' so that $m_0=n$ and $m_{k+1}=1$. 
Off diagonal elements of a $k$-step {\sc rsb} matrix $Q_{ab}$ for which 
$a$ and $b$ belong to the same ``$l$-family'' are labeled $q_l$. 
A 2-step {\sc rsb} 
matrix $\bf Q$ is then characterized (at zero magnetic field, that is, $q_0=0$) by 
two off-diagonal elements, $q_1$ and $q_2$, and two break points, 
$m_1$ and $m_2$,  
such that, in the $n\to 0$ limit, $0\le m_1\le m_2\le 1$. 
The only terms in the free-energy 
density that depend on these parameters are given by
\begin{eqnarray}
-{1\over 2\beta}\left[\log(\tilde{q}_d(0)/\beta - \langle\langle q \rangle\rangle )+{1\over m_1}
\log\left(1+m_1{q_1\over \tilde{q}_d(0)/\beta- \langle\langle q \rangle\rangle }\right)\right.
& & 
\nonumber\\
+\left.{m_2-1\over m_2}\log\left({\tilde{q}_d(0)/\beta -q_2\over
\tilde{q}_d(0)/\beta - \langle\langle q \rangle\rangle }\right)\right]
-{\beta\over 4}(m_2-1)q_2^p - {\beta\over 4}(m_1-m_2)q_1^p
\; ,
& &
\end{eqnarray}
where $ \langle\langle q \rangle\rangle =q_2+m_2(q_1-q_2)$.  
Hence, the extremization equations in order to $q_1$, $q_2$, 
$m_1$ and $m_2$ are given by
\begin{equation}
-{q_1\over
[\tilde{q}_d(0)+(m_2-1)\beta q_2+(m_1-m_2)\beta q_1]
(\tilde{q}_d(0)-\beta  \langle\langle q \rangle\rangle )}
+{p\over 2}q_1^{p-1} = 0 \; ,
\label{eq:aq1}
\end{equation}
\begin{eqnarray}
& &\left[1-{\beta q_1\over (
\tilde{q}_d(0)+(m_2-1)\beta q_2+(m_1-m_2)\beta q_1)}+
{\beta q_1-\tilde{q}_d(0)\over \tilde{q}_d(0)-\beta q_2}\right]
\nonumber\\
& & 
\;\;\;\;\;\;\;\;\;\;\;\;\;\;\;\;\;\;
\times {\beta\over \tilde{q}_d(0) -\beta  \langle\langle q \rangle\rangle }
+{p\over 2}\beta^2 q_2^{p-1}=0
\; ,
\label{eq:aq2}
\end{eqnarray}
\begin{eqnarray}
& &{1\over m_1^2}\log\left(
{{\tilde{q}}_d(0)-\beta  \langle\langle q \rangle\rangle \over
{\tilde{q}}_d(0)-(1-m_2)\beta q_2
+(m_1-m_2)\beta q_1}\right)+
{\beta^2\over 2}q_1^p\nonumber\\
& & 
\;\;\;\;\;\;\;\;\;\;\;\;\;\;\;\;\;\;\;
+{1\over m_1}{\beta q_1\over
{\tilde{q}}_d(0)-(1-m_2)\beta q_2
+(m_1-m_2)\beta q_1}=0
\label{eq:am1}
\end{eqnarray}
and
\begin{eqnarray}
& &{1\over m_2^2}\log\left(
{{\tilde{q}}_d(0)-\beta  \langle\langle q \rangle\rangle \over
{\tilde{q}}_d(0)-\beta q_2
}\right)+
{\beta^2\over 2}(q_1^p-q_2^p)\nonumber\\
& & 
\;\;\;\;\;\;\;\;\;\;\;\;\;
+{q_1-q_2\over \tilde{q}_d(0)-\beta  \langle\langle q \rangle\rangle }\left(
{\beta^2 q_1\over
{\tilde{q}}_d(0)-\beta  \langle\langle q \rangle\rangle 
+m_1\beta q_1}+{\beta\over m_2}\right)=0 \;, 
\label{eq:am2}
\end{eqnarray}
respectively. As can be easily seen, for $q_1=q_2$  
and $m_1=m_2$ (or $m_2=1$), the above set of equations  
yield the 1-step result. In order to demonstrate that the 1-step solution  
is unique,   
we now compute the linear 
combinations $-p \times $ Eq.~(\ref{eq:am1})$ + \beta^2q_1 \times $ 
Eq.~(\ref{eq:aq1}) and 
$\beta^2q_1\times $ Eq.~(\ref{eq:aq1}) $-q_2\times 
$ Eq.~(\ref{eq:aq2}) $-p\times $ Eq.~(\ref{eq:am2}) and 
define $x_p=(\beta q_1)/\tilde{q}_d(0)$ and 
$y=(\beta q_2)/\tilde{q}_d(0)$, to obtain the following equations:
\begin{eqnarray}\nonumber
& &{ x_p^2\over (1+(m_2-1)y_p+(m_1-m_2)x_p)(1-y_p-m_2(x_p-y_p))}\\
&+&{1\over m_1}{px_p\over 1+(m_2-1)y_p+(m_1-m_2)x_p}\\
\nonumber
&-&{p\over m_1^2}\log\left({1+(m_2-1)y_p+(m_1-m_2)x_p\over 1-y_p-m_2(x_p-y_p)}\right)=0
\label{eq:xy1}
\end{eqnarray}
and
\begin{eqnarray}
& &{x_p-y_p\over 1-y_p-m_2(x_p-y_p)}\left[{x_p(p+1)\over 1+(m_2-1)y_p+(m_1-m_2)x_p}+
{y_p\over 1-y_p}+{p\over m_2}\right]\nonumber\\
& &
\;\;\;\;\;\;\;\;\;\;\;\;\;\;\;\;\;\;\;\;
+{p\over m_2^2}\log\left({1-y_p-m_2(x_p-y_p)\over 1-y_p}\right)=0,
\label{eq:xy2}
\end{eqnarray}
respectively. 
At this point, it is useful to define the parameters
\begin{equation}
u_p={m_2(x_p-y_p)\over 1-y_p}
\end{equation}
and
\begin{equation}
v_p={m_1x_p\over 1-y_p}
\end{equation}
and rewrite eqs. (\ref{eq:xy1}) and (\ref{eq:xy2}) as
\begin{equation}
{v_p^2\over (1-u_p+v_p)(1-u_p)}+{v_p p\over 1-u_p+v_p}
-p\log\left(1+{v_p\over 1-u_p}\right)=0 
\label{eq:vp}
\end{equation}
and
\begin{equation}
\left({m_2\over m_1}{v_p(1+p)\over 1-u_p+v_p}-u_p+{m_2\over m_1}
v_p+p\right){u_p\over 1-u_p}
-p\log\left({1\over 1-u_p}\right)=0,
\label{eq:up}
\end{equation}
respectively. 
The only real solution to the above equations is $u_p=0$ 
and $v_p$ equals its 1-step value (for example, $v_p=1.81696$  
for $p=3$). 
Therefore, we can conclude that the 2-step {\sc rsb} 
scheme yields back the 1-step one.


\begin{thebibliography}{99}
\bibitem{footnote}
More precisely,this connection allows us  
to determine how the  real-time correlation function $C(t,t')$ 
approaches its plateau value $q_{\sc ea}$ for $(t-t') \ll t'$ but not how it
decays below $q_{\sc ea}$ for  $(t-t') \gg t'$.
\bibitem{footnote2}
In Ref.~\cite{leti1} a discontinuous transition with $X=1$ all 
along the critical line was {\it assumed} and scaling laws for the 
critical behaviour derived on that basis. The results of this paper
show that this assumption is not applicable to the model.
\end{thebibliography}
\end{document}